\documentclass[aip,preprint,showpacs,superscriptaddress]{revtex4-1}

\usepackage{amsmath,graphics}
\usepackage[next]{inputenc}
\usepackage[dvips]{epsfig}
\usepackage[colorlinks=true,citecolor=blue,linkcolor=blue]{hyperref}
\usepackage{bbm}
\usepackage{bbold}
\usepackage{booktabs}
\usepackage{multirow}
\usepackage{hhline}
\usepackage{graphicx}
\usepackage{amsmath}
\usepackage{multirow}
\usepackage[usenames, dvipsnames]{color}
\usepackage{color,soul}
\usepackage{xcolor}
\usepackage{latexsym}

\usepackage{subfigure}
\usepackage{hyperref}
\usepackage{float}
\usepackage{amssymb}

\begin{document}
\preprint{AIP/123-QED}

\title{Magnetic field-induced non-trivial electronic topology in Fe$_{3-x}$GeTe$_2$}

\author{Juan Macy$^\ddag$}
\affiliation{National High Magnetic Field Laboratory, Florida State University, Tallahassee, Florida 32310, USA}
\affiliation{Department of Physics, Florida State University, Tallahassee, Florida 32306, USA}
\thanks{These authors contributed equally to this work}

\author{Danilo Ratkovski$^\ddag$}
\affiliation{National High Magnetic Field Laboratory, Florida State University, Tallahassee, Florida 32310, USA}
\thanks{These authors contributed equally to this work}

\author{Purnima P. Balakrishnan$^\ddag$}
\affiliation{NIST Center for Neutron Research, NIST, Gaithersburg, MD 20899, USA}

\author{Mara Strungaru}
\affiliation{Institute for Condensed Matter Physics and Complex Systems, School of Physics and Astronomy,The University of Edinburgh, EH9 3FD, UK}

\author{Yu-Che Chiu}
\affiliation{National High Magnetic Field Laboratory, Florida State University, Tallahassee, Florida 32310, USA}
\affiliation{Department of Physics, Florida State University, Tallahassee, Florida 32306, USA}

\author{Aikaterini Flessa Savvidou}
\affiliation{National High Magnetic Field Laboratory, Florida State University, Tallahassee, Florida 32310, USA}
\affiliation{Department of Physics, Florida State University, Tallahassee, Florida 32306, USA}

\author{Alex Moon}
\affiliation{National High Magnetic Field Laboratory, Florida State University, Tallahassee, Florida 32310, USA}
\affiliation{Department of Physics, Florida State University, Tallahassee, Florida 32306, USA}

\author{Wenkai Zheng}
\affiliation{National High Magnetic Field Laboratory, Florida State University, Tallahassee, Florida 32310, USA}
\affiliation{Department of Physics, Florida State University, Tallahassee, Florida 32306, USA}

\author{Ashley Weiland}
\affiliation{Department of Chemistry and Biochemistry, The University of Texas at Dallas, Richardson, Texas 75080, USA}

\author{Gregory T. McCandless}
\affiliation{Department of Chemistry and Biochemistry, The University of Texas at Dallas, Richardson, Texas 75080, USA}

\author{Julia Y. Chan}
\affiliation{Department of Chemistry and Biochemistry, The University of Texas at Dallas, Richardson, Texas 75080, USA}

\author{Govind S. Kumar}
\affiliation{Department of Chemistry \& Biochemistry, Florida State University, 95 Chieftan Way, Tallahassee, FL 32306, USA}

\author{Michael Shatruk}
\affiliation{Department of Chemistry \& Biochemistry, Florida State University, 95 Chieftan Way, Tallahassee, FL 32306, USA}

\author{Alexander J. Grutter}
\affiliation{NIST Center for Neutron Research, NIST, Gaithersburg, MD 20899, USA}

\author{Julie A. Borchers}
\affiliation{NIST Center for Neutron Research, NIST, Gaithersburg, MD 20899, USA}

\author{William D. Ratcliff}
\affiliation{NIST Center for Neutron Research, NIST, Gaithersburg, MD 20899, USA}

\author{Eun Sang Choi}\email{echoi@magnet.fsu.edu}
\affiliation{National High Magnetic Field Laboratory, Florida State University, Tallahassee, Florida 32310, USA}
\affiliation{Department of Physics, Florida State University, Tallahassee, Florida 32306, USA}

\author{Elton J. G. Santos}\email{esantos@exseed.ed.ac.uk}
\affiliation{Institute for Condensed Matter Physics and Complex Systems, School of Physics and Astronomy,The University of Edinburgh, EH9 3FD, UK}
\affiliation{Higgs Centre for Theoretical Physics, The University of Edinburgh, EH9 3FD, UK}

\author{Luis Balicas}\email{balicas@magnet.fsu.edu}
\affiliation{National High Magnetic Field Laboratory, Florida State University, Tallahassee, Florida 32310, USA}
\affiliation{Department of Physics, Florida State University, Tallahassee, Florida 32306, USA}

\begin{abstract}
The anomalous Hall, Nernst, and thermal Hall coefficients of the itinerant ferromagnet Fe$_{3-x}$GeTe$_2$ display anomalies upon cooling that are consistent with a topological transition that could induce deviations with respect to the Wiedemann-Franz (WF) law.  This law has not yet been validated for the anomalous transport variables, with recent experimental studies yielding material-dependent results. Nevertheless, the anomalous Hall and thermal Hall coefficients of Fe$_{3-x}$GeTe$_2$ are found, within our experimental accuracy, to satisfy the WF law for magnetic-fields $\mu_0H$ applied along its \emph{c}-axis. Remarkably, large anomalous transport is also observed for $\mu_0H \| a$-axis with the field aligned along the gradient of the chemical potential generated by thermal gradients or electrical currents, a configuration that should not lead to their observation. These anomalous planar quantities are found to not scale with the component of the planar magnetization ($M_{\|}$), showing instead a sharp decrease beyond $\mu_0 H_{\|} = $ 4 T or the field required to align the magnetic moments along $\mu_0 H_{\|}$. We argue that chiral spin structures associated with Bloch domain walls lead to a field dependent spin-chirality that produces a novel type of topological transport in the absence of interaction between the magnetic field and electrical or thermal currents. Locally chiral spin-structures are captured by our Monte-Carlo simulations incorporating small Dzyaloshinskii-Moriya and biquadratic exchange interactions. These observations reveal not only a new way to detect and expose topological excitations, but also a new configuration for heat conversion that expands the current technological horizon for thermoelectric energy applications.
\end{abstract}

\date{\today}

\maketitle

\section{Introduction}

\begin{figure*}[ht]
\begin{center}
	\includegraphics[width = 14 cm]{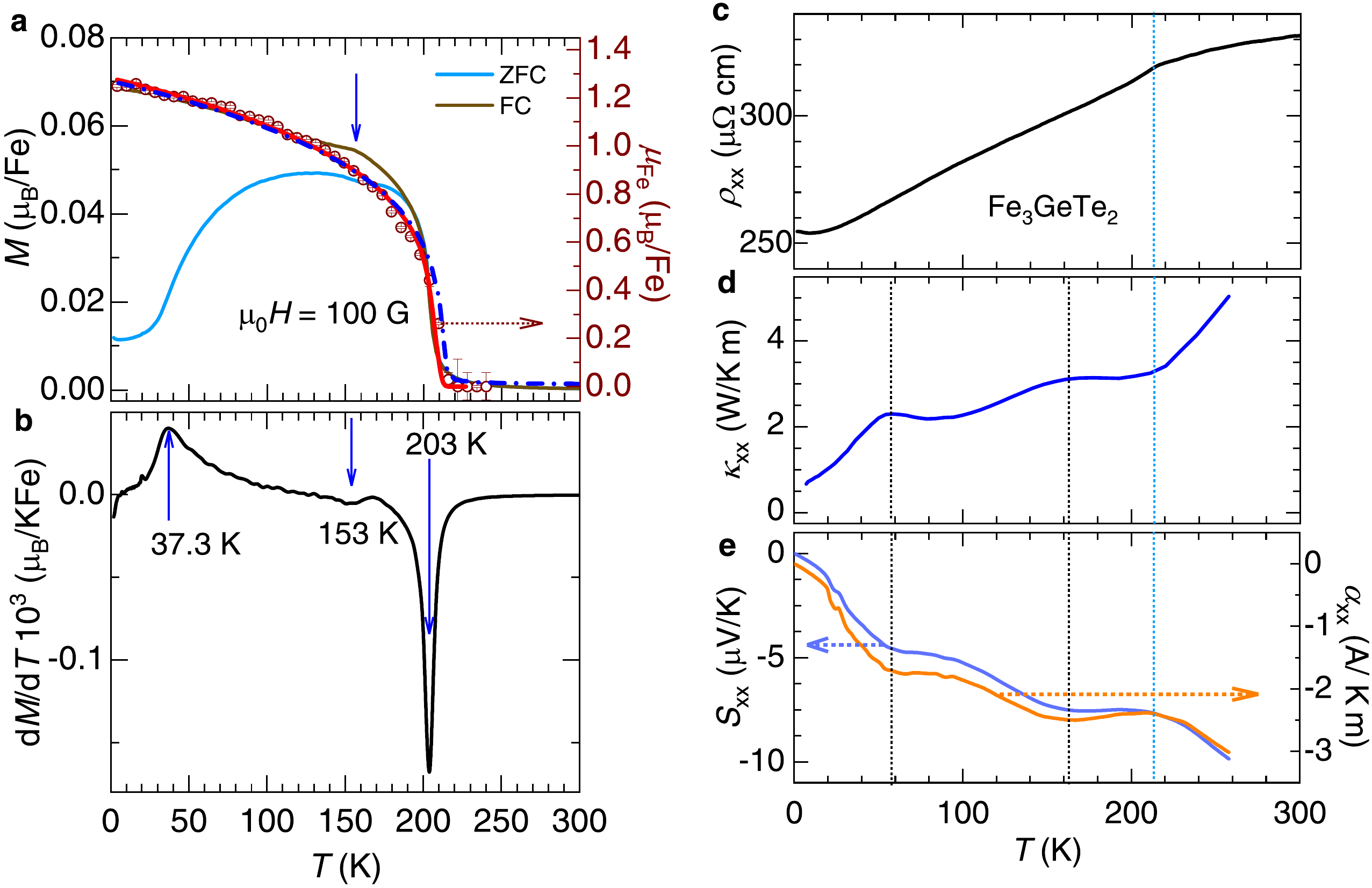}
	\caption{\textbf{Magnetization and magnetic moment according to neutron scattering as a function of the temperature for Fe$_{3-x}$GeTe$_2$}.
	{\bf a,} Magnetization $M$ ($\mu_B$ per Fe atom) for a \textcolor{red}{Fe$_{3.05}$GeTe$_2$} single-crystal subjected to $\mu_0 H = 100$ G applied along its inter-layer direction as a function of the temperature $T$. Here, error bars represent one standard deviation. Faint blue and brown curves correspond to zero-field and field-cooled conditions, respectively. Brown markers represent the normalized magnetic moment per Fe atom (e.g.  $\mu_{Fe} \propto \sqrt{z}$ as a function of $T$,  where $z$ is the intensity of the neutron diffraction peak scattered by the inter-planar magnetic order).  The structural contribution to the neutron scattered intensity has been subtracted.  Red line is a fit to $[I=I_0(1-T/T_c)^{\beta}]^{1/2}$,  where $I$ is the scattered intensity,  $I_0$ is its value in the $T \rightarrow 0$ K limit, $T_c = (207.0 \pm 0.3)$ K the Curie temperature.  This data was collected under $\mu_0 H = 0$ T. 	Dashed blue line represents the calculated $\mu_{Fe}(T)$ according to the spin Hamiltonian in Eq. \ref{gen_ham}, which includes biquadratic exchange interactions as described in the main text. {\bf b,} $\partial M/ \partial T$ as a function of $T$ for the zero-field cooled trace. Arrows indicate anomalies including the Curie temperature at $\approx$ 203 K (middle point of the transition). Additional anomalies are seen at lower $T$s,  particularly near $ \approx$ 150 K  and 50 K (see SI file for data from yet another sample confirming these anomalies). {\bf c,} Resistivity $\rho$ as a function of $T$ for a \textcolor{red}{Fe$_{2.90}$GeTe$_2$} crystal, where a sharp change in slope is observed at $T_c \simeq 212$ K (indicated by blue dotted line). {\bf d,} Thermal conductivity $\kappa_{xx}$, and (e), thermoelectric power $S_{xx}$ as well as Peltier conductivity $\alpha_{xx} = \sigma_{xx}S_{xx}$ as functions of $T$, respectively. Notice the broad anomalies centered around 60 K and 160 K (indicated by black dotted lines). These lower $T$ anomalies contrast with the lack of any anomaly in the neutron data.} \label{diagonal_components}
\end{center}
\end{figure*}

\begin{figure*}[htb]
\begin{center}
	\includegraphics[width=15 cm]{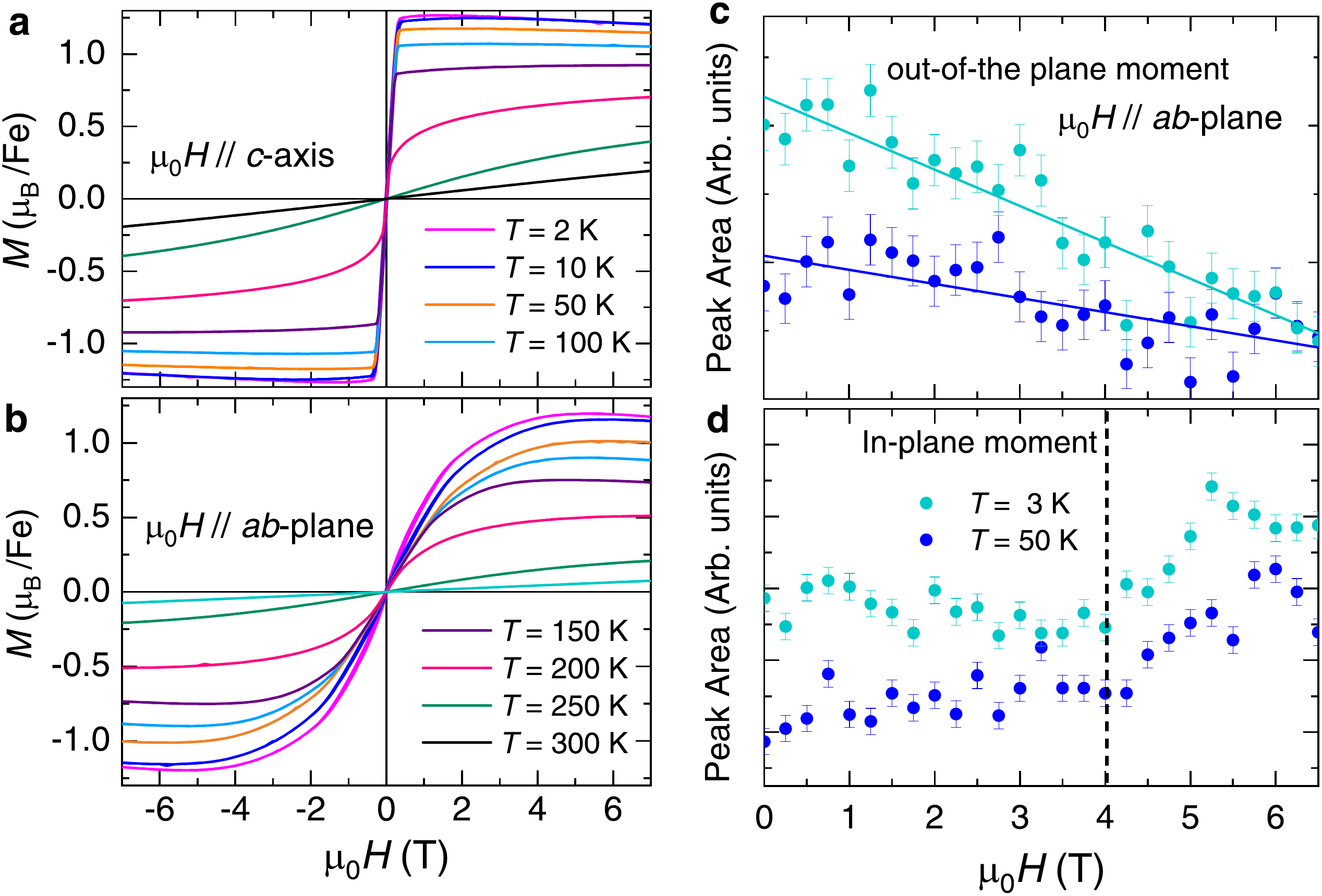}
	\caption{\textbf{Magnetization and neutron scattering intensity as functions of magnetic field.}
	{\bf a-b, } Magnetization $M$ as function of the magnetic field $\mu_0H$ for several temperatures $T$,
	for two orientations of a \textcolor{red}{Fe$_{3.05}$GeTe$_2$}  single-crystal with fields along the $c-$axis and $ab-$plane,  respectively.
	{\bf c-d,} Integrated area of neutron diffraction peaks respectively for the (002) peak, which is sensitive with respect to the in-plane moments,  and the (100) peak, which is sensitive to the moments aligned along the inter-planar direction. Error bars represent one standard deviation.  $\mu_0H$ is applied along the planar direction for $T = 3$ K and 50 K.  These peaks contain both the magnetic and structural contributions to the neutron scattering.  Notice that their area,  which is monotonic with the component of the magnetic moment, progressively decreases along the out-of-plane direction while it continuously increase along the planar one, in agreement with the magnetization.  Lines, or linear fits, act as guides to the eyes.} 
\end{center}
\end{figure*}

\begin{figure*}[htb]
	\begin{center}
		\includegraphics[width=1\columnwidth]{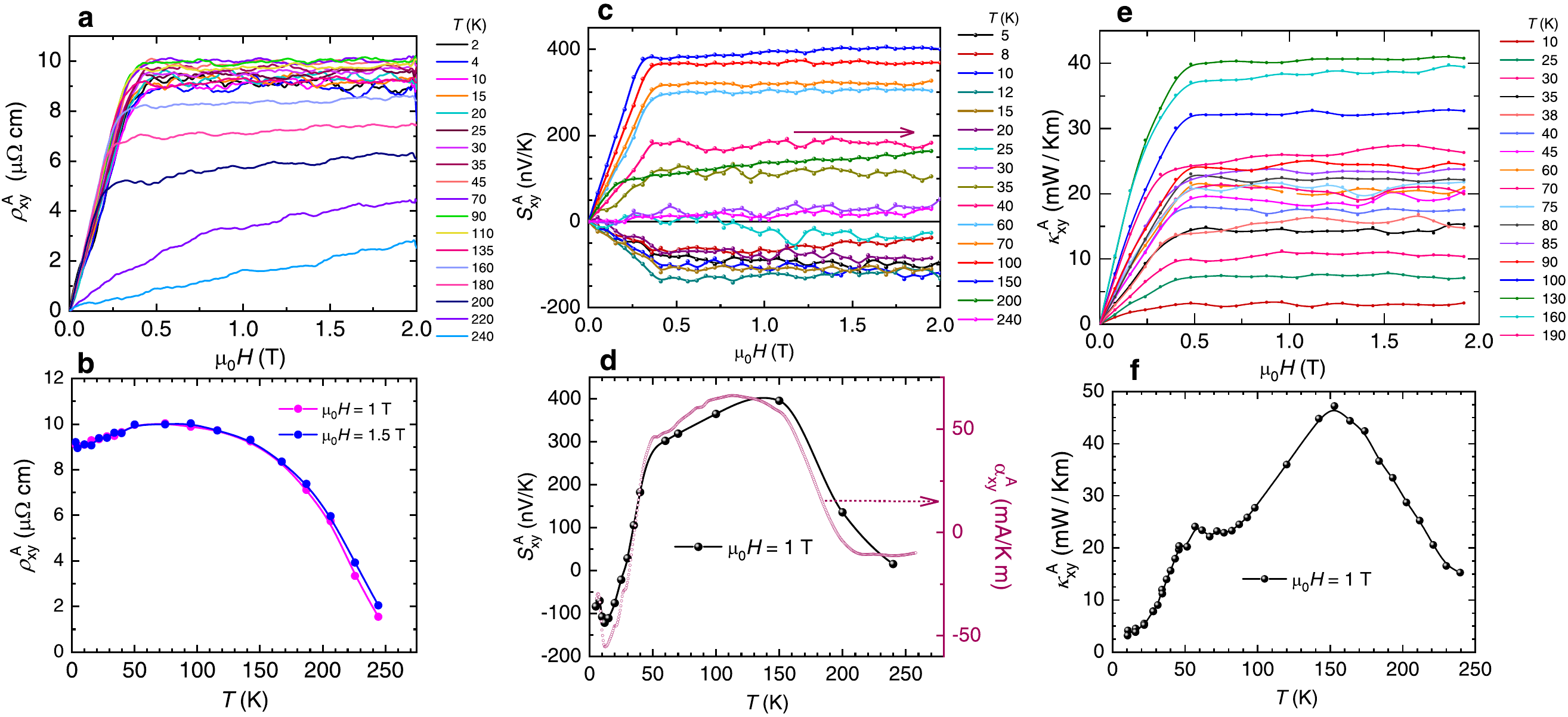}
		\caption{\textbf{Anomalous transport variables for magnetic fields along the inter-layer direction.}
		{\bf a,} Anomalous Hall resistivity $\rho_{xy}^A = R_H \cdot t$,  where $R_H=V_H/I$ is the Hall resistance,  $t$ the sample thickness and $I$ the electrical current respectively,  for a Fe$_{3-x}$GeTe$_2$ crystal \textcolor{red}{$(x \simeq 0.15)$} as a function of $\mu_0H \|$ \emph{c}-axis and for several $T$s.  We have not subtracted the comparatively small conventional Hall signal which would add a slope superimposed onto the Hall like plateau. {\bf b,} $\rho_{xy}^A$ under $\mu_0H = 1$ T (magenta) and 1.5 T (blue) as a function of $T$. In contrast to $M$ (Fig. 1a), $\rho_{xy}^A$ decreases slightly below $T \leq 50$ K. {\bf c,} Anomalous Nernst effect $S_{xy}^A$ as a function of $\mu_0H$ for several $T$s. (d) $S_{xy}^A$ as a function $T$ collected under $\mu_0H=1$ T applied along the \emph{c}-axis. Notice how $S_{xy}^A$ changes sign upon cooling below $\leq 50$ K after reaching a maximum at $\approx 150$ K. {\bf e,} Anomalous thermal Hall conductivity $\kappa_{xy}^A$ as a function of $\mu_0H \|$ \emph{c}-axis for several \emph{T}s. \textbf{f,} $\kappa_{xy}^A$ under $\mu_0H = 1$ T as a function of $T$. Both $\kappa_{xy}^A$ and $S_{xy}^A$ display anomalies at $\approx 50$ and $\approx 150$ K. These data were collected on the same single-crystal using the same electrodes, while Nernst and thermal Hall were measured simultaneously.} \label{c_axis_anomalous}
	\end{center}
\end{figure*}

\begin{figure}[htb]
	\begin{center}
		\includegraphics[width = 7 cm]{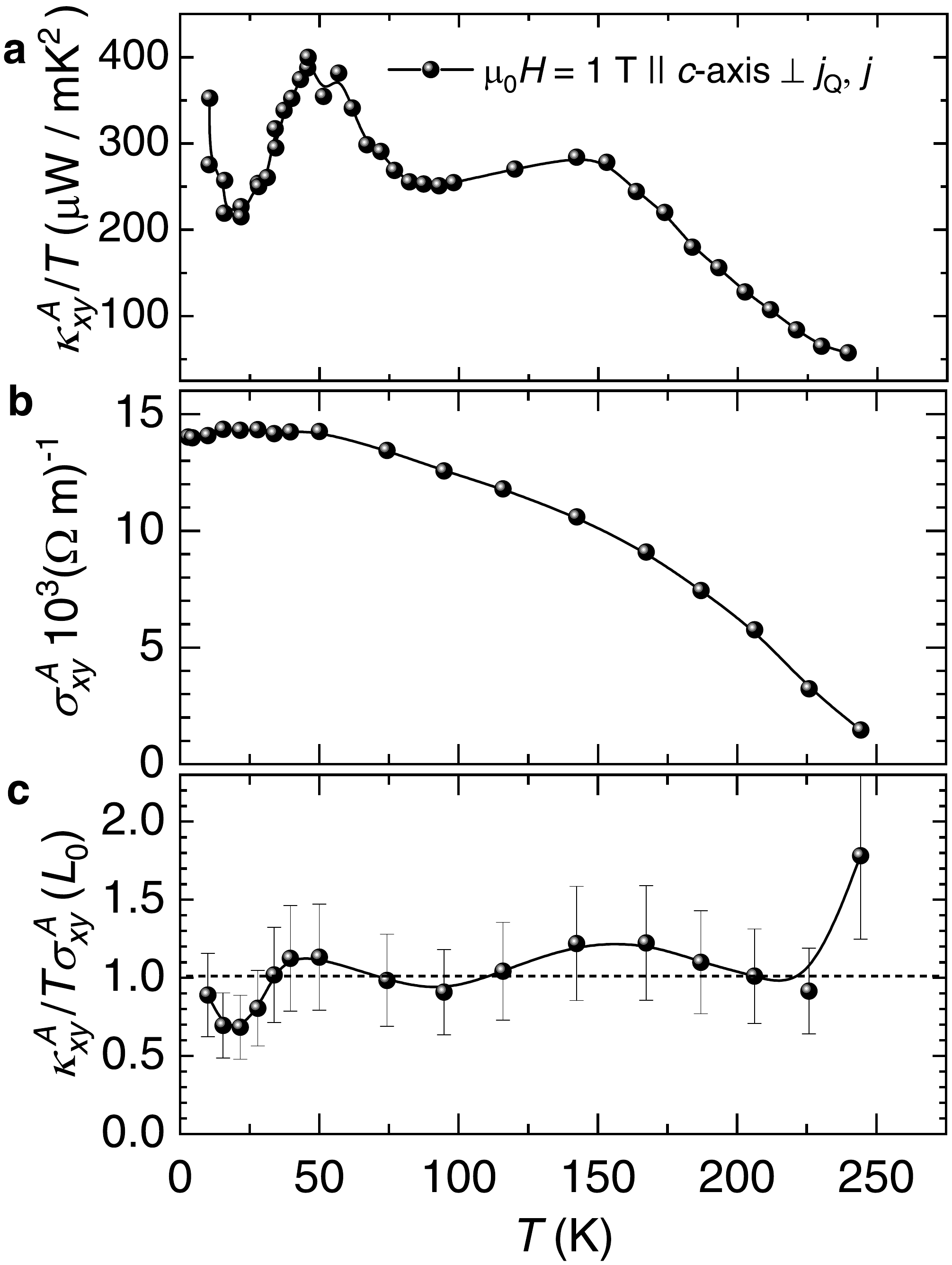}
		\caption{\textbf{Anomalous transport variables of Fe$_{3-x}$GeTe$_2$ and the Wiedemann-Franz law.}
		{\bf a, } Anomalous thermal Hall conductivity $\kappa_{xy}^A$ normalized by $T$ and measured under $\mu_0H=$ 1 T $\|$ \emph{c}-axis as a function of $T$.
		{\bf b,} Absolute value of the anomalous Hall conductivity $\sigma_{xy}^A$ collected under $\mu_0H=$ 1 T $\|$ \emph{c}-axis as a function of $T$.
		{\bf c,} $\kappa_{xy}^A/T\sigma_{xy}^A$ in units of the Lorentz number $L_0 = 2.44 \times 10^{-8}$ W$\Omega$K$^{-2}$. The anomalous transport quantities in Fe$_{3-x}$GeTe$_2$ would satisfy the Wiedemann-Franz law if one included error bars defined by the thermal and electrical noise intrinsic to our experimental set-up.} \label{WF}
	\end{center}
\end{figure}

\begin{figure*}[htb]
	\begin{center}
		\includegraphics[width = 16 cm]{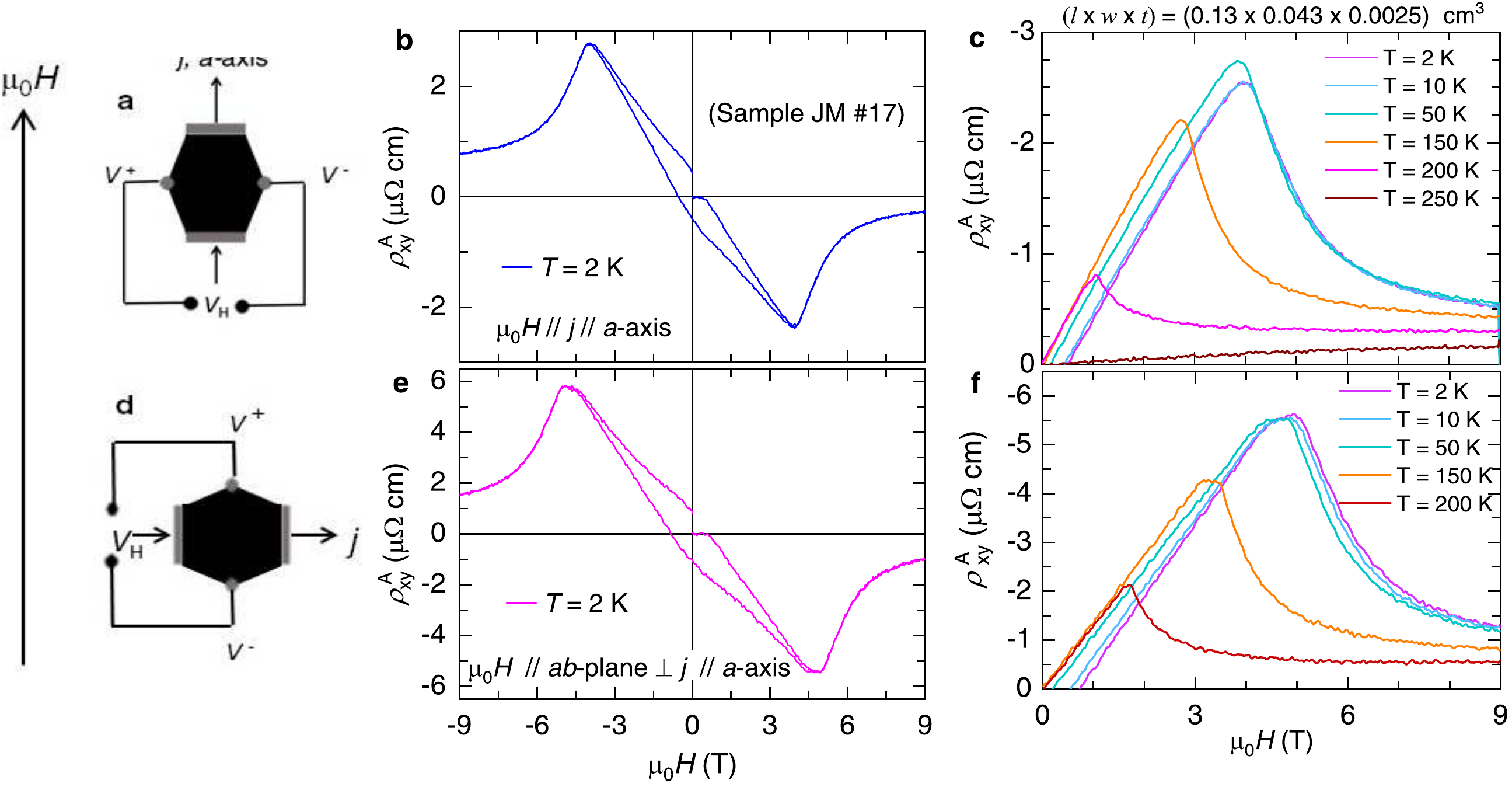}
		\caption{\textbf{Hall like effect for unconventional magnetic field orientations.}
		{\bf a,} Configuration of electrical contacts and relative orientation
		between the external magnetic field $\mu_0H$ and the electrical current $j$ (with both quantities aligned) used for the measurements shown in {\bf b-c}.
		{\bf b,} Hall like voltage normalized by the
		electrical current and multiplied by the sample thickness
		along the out of the plane direction,  which yields a rather
		anomalous and asymmetric Hall like resistivity $\rho_{xy}^A$.
		Notice the hysteresis seen around zero field at a temperature $T = 2$ K.
		{bf c,} Antisymmetric component of the Hall like resistivity,
		i.e. $(\rho_{xy}^A (+ \mu_0 H) - \rho_{xy}^A (- \mu_0 H))/2$.
		Hysteresis leads to the finite intercept with the magnetic field
		axis seen at low $T$s.
		{\bf d,} Configuration of measurements for $\mu_0H$
		along a planar direction but perpendicular to $j$ which
		was used for the measurements shown in {\bf e-f}.
		{\bf e} $\rho_{xy}^A$ as a function of $\mu_0H$ indicating
		that it is perfectly asymmetric with respect to both field orientations.
		{\bf f,} Anti-symmetrized $\rho_{xy}^A$.  This data was collected
		from crystal JM17 (nominally, Fe$_{2.84}$GeTe$_2$).}
	\end{center}
\end{figure*}

\begin{figure*}[ht]
	\centering
	\includegraphics[width=1\columnwidth]{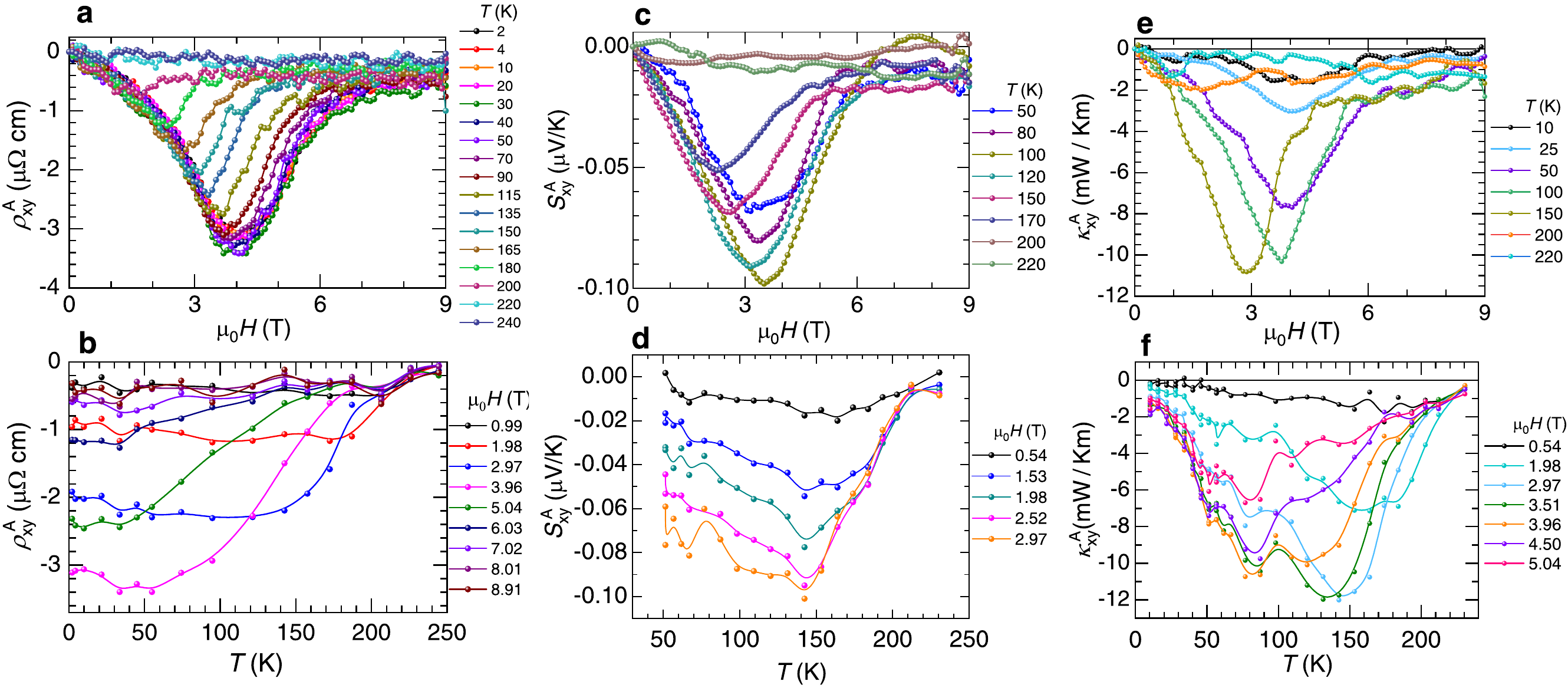}
	\caption{\textbf{Field-induced maxima in the anomalous transport variables for currents and thermal gradients aligned along $\mu_0H$ $\|$ to the a-axis.}
	{\bf a,} Anomalous Hall resistivity $\rho_{xy}^A$ for a
	Fe$_{2.84}$GeTe$_2$ crystal as a function of $\mu_0H$
	 applied along the \emph{a}-axis and for several $T$s.
	 $\rho_{xy}^A$ displays a peak as a function of $\mu_0H$
	 whose position is $T$-dependent.
	 {\bf b,} $\rho_{xy}^A$ as a function of $T$ for several
	 values of $\mu_0H$ applied along a planar direction.
	 {\bf c,} Anomalous Nernst effect $S_{xy}^A$ as a function
	 of $\mu_0H$ along a planar direction for several $T$s.
	 {\bf d,} $S_{xy}^A$ as a function $T$ collected under
	 several values of the field applied along a planar direction.
	 {\bf e,} Anomalous thermal Hall conductivity $\kappa_{xy}^A$
	 as a function of $\mu_0H$ applied along a planar
	 direction and for several values of \emph{T}.
	 {\bf f,} $\kappa_{xy}^A$ as a function of $T$ and for several
	 values of $\mu_0H$ applied along a planar direction.
	 The anomalous transport variables follow a similar
	 dependence on magnetic field which,
	 contrasts with the one followed by the magnetization.}
	 \label{ab-plane_anomalous}
\end{figure*}

\begin{figure}[htb]
	\begin{center}
		\includegraphics[width = 8 cm]{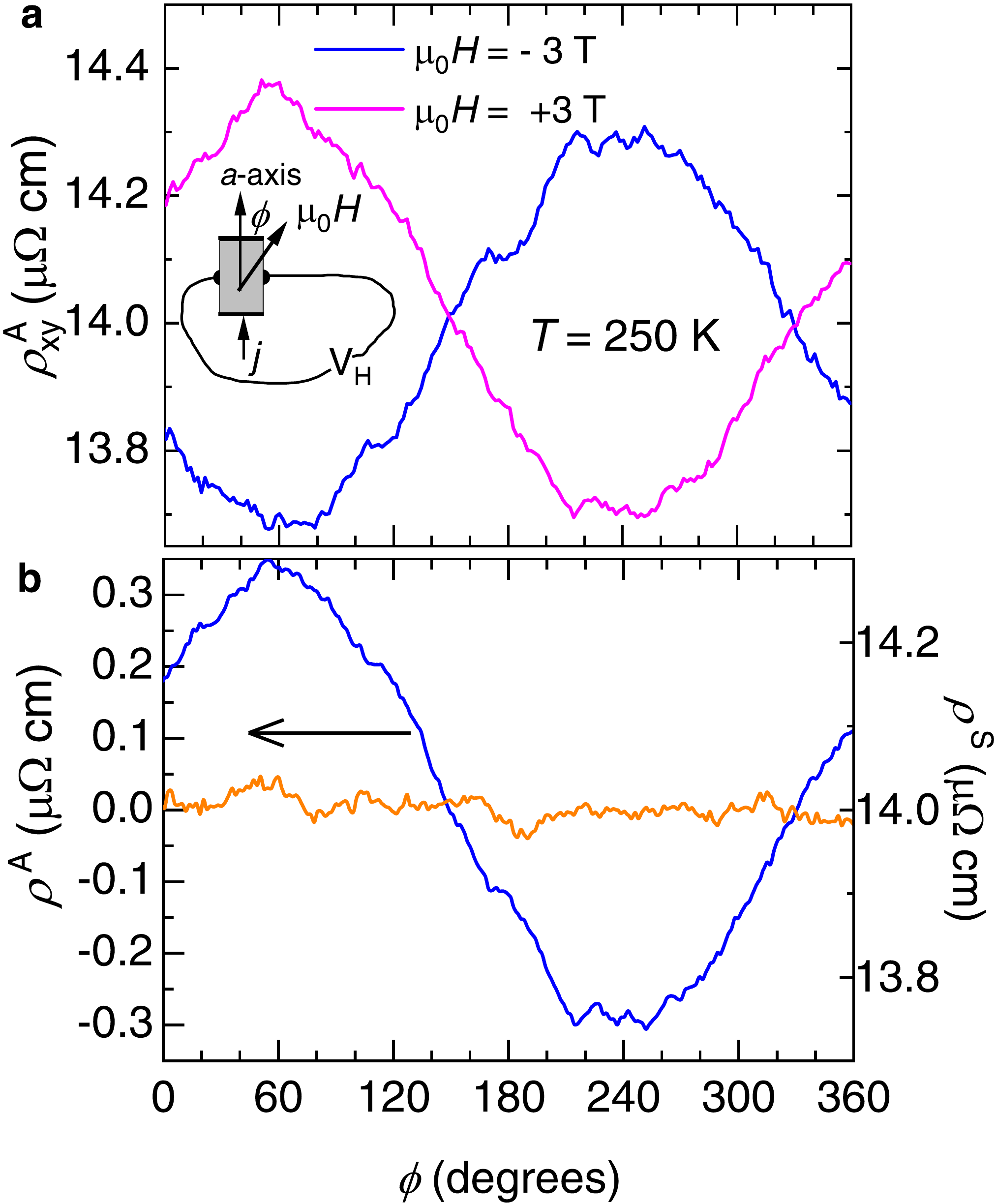}
		\caption{\textbf{Hall like signal in the paramagnetic state of Fe$_{3-x}$GeTe$_2$ for fields rotating within the conducting planes.}
		{\bf a,} Raw Hall-like signal for magnetics fields
		rotating between a direction along the current density $j$
		and a direction perpendicular to it but still along the planar
		direction.  Angle $\phi$ defines the angle between $\mu_0H$ and $j$.
		Magenta trace corresponds to $\mu_0H = + 3$ T,
		while blue trace corresponds to $\mu_0H = - 3$ T.
		{\bf b,}  Antisymmetric component $\rho^{\text{A}}$
		superimposed on both traces (in blue),
		and their average $\rho^{\text{S}}$  (orange trace).
		Even the paramagnetic phase displays an anomalous Hall like
		response in a geometry where no Hall signal should be observed.}
	\end{center}
\end{figure}

The current interest in topologically non-trivial compounds is in the promise of observing novel magneto-opto-electronic phenomena with potential technological relevance, hitherto unobserved in conventional materials. Examples include the generation of photocurrents with circularly polarized light \cite{Ma1,Orenstein}, the observation of a Hall like signal in the absence of broken time-reversal symmetry in Weyl semimetals \cite{Ma2}, or the observation of large anomalous Hall and Nernst-effects at room temperature in a non-collinear antiferromagnet displaying magnetic Weyl fermions \cite{Satoru, Ikhlas, Kuroda}. For instance, the large anomalous Nernst effect observed at room temperature in magnetic Weyl semimetals has been proposed as an effective alternative to thermoelectric energy conversion \cite{Akito, Ikhlas}

In this context, Fe$_{3-x}$GeTe$_2$ is a layered, van der Waals like ferromagnet displaying a simple collinear spin arrangement with the magnetic moments oriented along the out-of-the plane direction \cite{Deiseroth, May1}. However, this simple magnetic order is claimed to trigger complex phenomena such as i) the development of a Kondo lattice below a coherence temperature of $\approx 150$ K \cite{Kondo}, ii) electric field tuning of its Curie temperature $T_c= (220 \pm 10)$ K up to room temperature \cite{Deng}, iii) skyrmions  \cite{Ding, Pei, skyrmion3}, iv) and very large anomalous Hall and Nernst coefficients claimed to result from its non-trivial electronic topology \cite{Kim, Nernst}.

Fe$_{3-x}$GeTe$_2$ displays a comparatively high Curie temperatures $T_c$, relative to the magnetic ordering temperature of other two-dimensional magnetic systems, that is  ranging from 150 to 220 K depending on the Fe occupancy \cite{May1, Abrikosov, Deiseroth, Liu}. Fe$_{3-x}$GeTe$_2$ can be understood as containing van der Waals (vdW) bonded Fe$_{3-x}$GeTe$_2$ slabs or, as discussed in Ref. \cite{Seo}, as a scaffold with a lattice akin to that of the transition metal dichalcogenides but stuffed with Fe atoms. Its structure leads to two inequivalent Fe sites, Fe$_I^{3+}$ and Fe$_{II}^{2+}$, within the Fe$_{3-x}$GeTe$_2$ slab \cite{Deng, Deiseroth}. Partially filled Fe-\emph{d} orbitals dominate the band structure around the Fermi level producing itinerant ferromagnetism in bulk Fe$_{3-x}$GeTe$_2$ \cite{Gong}. As a result of the reduced crystallographic symmetry inherent to its layered structure (space group 194; $P6_3/mmc$), bulk Fe$_{3-x}$GeTe$_2$ exhibits a strong magneto-crystalline anisotropy \cite{Zhuang}.
The observation of skyrmions \cite{Ding, Pei, skyrmion3} and spin spirals on its surface \cite{Fe312_DMI} can only be reconciled with a sizeable Dzyaloshinskii-Moriya interaction, whose origin is discussed in Ref. \cite{Laref}, although inelastic neutron scattering would suggest negligible inter-layer exchange interactions but a prominent role for the single ion anisotropy \cite{Calder}.

Fe$_{3-x}$GeTe$_2$ is claimed to correspond to a rare example of a ferromagnetic topological nodal line semimetallic system for which electronic correlations are claimed to be relevant \cite{Kim, ZhuJ}. In particular, the existence of a gapped Dirac nodal line, and its remanent Berry curvature, would explain the very large anomalous Hall and Nernst coefficients in Fe$_{3-x}$GeTe$_2$ with concomitantly large anomalous Hall and Nernst angles \cite{Kim, Nernst}.  Therefore, it is an ideal system to explore the relation between the off-diagonal anomalous variables associated to charge (Hall-effect) and heat/entropy (thermal Hall-effect) transport since this is a relatively unexplored subject from both the theoretical and experimental perspectives \cite{Behnia}. Although the anomalous transport variables were found to satisfy the Wiedemann-Franz (WF) law in few compounds \cite{Behnia2}, Mn$_3$Ge is claimed to violate it due to a mismatch in thermal and electrical summations of the Berry curvature over the Fermi surface \cite{Behnia}. At first glance, Fe$_{3-x}$GeTe$_2$ offers a relatively simple magnetic system to test its general validity for the anomalous transport variables since previous assertions in favor of its violation (see, e.g. \cite{tanatar, wakeham}) were claimed to result from artifacts \cite{Behnia, Behnia2}.

This compound has a potential for spintronics applications through the electrical control of its magnetic domains and skyrmions.
Neutron scattering and magnetization measurements reveal a ground state with magnetic moments pointing collinearly along the c-axis \cite{May1, WangY} as confirmed by us.
This collinear spin arrangement leads to stripe like magnetic domains according to in-situ Lorentz transmission electron microscopy \cite{Wu, Ding, Pei}. Remarkably, application of a magnetic field along the \emph{c}-axis, induces the formation of magnetic bubbles or magnetic skyrmions as those domains with spins pointing against the field are suppressed by it \cite{Ding, Pei}. Skyrmions could result from the Dzyaloshinskii-Moriya interaction since the inequivalent Fe sites form a lattice that lacks inversion symmetry \cite{Laref}. The size of these domains are susceptible to manipulation via pulses of electrical current which apparently can also induce skyrmion bubbles \cite{Pei}. As Fe$_{3-x}$GeTe$_2$ still exhibits robust ferromagnetism with a strong perpendicular anisotropy even when mechanically exfoliated down to the monolayer limit \cite{Fei}, it has potential for 2D spintronics applications.

In this manuscript, we report on novel anomalous Hall, Nernst and thermal Hall effects in Fe$_{3-x}$GeTe$_2$ having clear topological origin associated to magnetic field-induced spin textures. Typically, the anomalous transport variables are observed in solids characterized by broken time-reversal symmetry (e.g. ferromagnets) result from the spin-orbit coupling and are conceptually treated in terms of the Berry phase \cite{AHE_review}. For magnetic fields $\mu_0 H$ along the inter-planar direction (\emph{c}-axis) and electrical currents flowing a long a planar direction (\emph{a}-axis), we confirm the observation of a very large anomalous Hall conductivity $\sigma_{xy}$ accompanied by a concomitantly large anomalous Nernst coefficient $S_{xy}^A$ and a very large anomalous thermal Hall signal $\kappa_{xy}^A$. In contrast to Ref. \onlinecite{Behnia}, we find that the anomalous variables in Fe$_{3-x}$GeTe$_2$ do satisfy the Wiedemann-Franz law over the range of temperatures measured, i.e. 2 to 225 K. However, the diagonal as well as the off-diagonal components of the thermal transport variables reveal anomalies around $\approx 150$ K, as well as around 50 K which corresponds to the onset of a change in the sign of the Nernst signal upon cooling. Concomitant anomalies are seen in the magnetization but not in the heat capacity or neutron scattering data, suggesting the possibility of a very small canting of the magnetic moments with respect to the inter-planar direction. Surprisingly, in Fe$_{3-x}$GeTe$_2$ we observe a novel type of anomalous transport variables even when the external magnetic field is aligned along the gradient of the chemical potential (or along any planar direction), e.g. implying a Hall-like signal in absence of Lorentz force. For this geometry, $\rho_{xy}^A$, $S_{xy}^A$, $\kappa_{xy}^A$ all increase as the in-plane field $\mu_0 H_{\|}$ increases, peaking around $\mu_0 H_{\|} \simeq 4$ T and then saturating beyond $\mu_0H_{\|} \simeq 6$ T, the field where the magnetization $M_{\|}$ is also observed to saturate. The size of the effect is comparable among samples having distinct geometries (e.g. different inter-layer thicknesses $t$). Therefore, we conclude that a field dependent chiral spin texture leads to a finite field-induced spin chirality that affects the Berry phase of the charge carriers leading to hitherto unreported anomalous Hall, Nernst and thermal Hall effects having a topological origin. Monte-Carlo simulations that incorporate small Dzyaloshinskii-Moriya and biquadratic exchange interactions yield spiral spin textures for the domain walls separating the serpentine domains with the emergence of skyrmions, and possibly also skyrmion tubes, upon application of an external magnetic field.  An asymmetric anomalous planar Hall-effect is also seen in the paramagnetic state suggesting that even in the absence of magnetic order Fe$_{3-x}$GeTe$_2$ bears non-trivial electronic topology.

\section{Results and discussion}
\subsection{Basic properties}
The onset of ferromagnetic order in a Fe$_{3.05}$GeTe$_2$ single-crystal at $T_c \approx 210$ K is observed as a sharp increase of the magnetization $M$ per Fe atom as a function of the temperature $T$ (Fig. 1a). A mild anomaly is seen in $M(T)$ for \emph{T}s between 150 K and 170 K, a range of temperatures claimed to coincide with the onset of a Kondo lattice \cite{Kondo} in this \emph{d}-electron system. Below 100 K, one observes a pronounced deviation between traces collected under zero-field (blue) and field cooled (magenta) conditions due to the development of striped or labyrinthine domains \cite{Wu, Ding, Pei} with their opposite magnetic moments oriented along the \emph{c}-axis. The same panel plots the $\mu_{\text{Fe}}$ extracted from the scattered intensity by the (100) magnetic Bragg peak (associated with the out of the plane magnetic moment) as a function of $T$. As seen, $\mu_{\text{FE}}$ decreases continuously as $T$ increases hence not providing evidence for additional magnetic phase-transitions; the Bragg peak at (002), which is sensitive to in-plane moment, remains constant through $T_c$ indicating the absence of an in-plane moment at zero-field. Neutron scattering supports previous reports indicating a collinear ferromagnetic ground state with moments along the \emph{c}-axis \cite{May1} although it leads to skyrmions \cite{Wu,Ding,skyrmion3} and chiral spin spirals at its surface \cite{Fe312_DMI}. A more detailed neutron scattering study exploring a broader region in \emph{k}-space is required to reconcile these contrasting observations. The derivative of $M$ as a function of $T$ (Fig. 1b) highlights the magnetization anomalies, corresponding to the middle point $T_c$ as well as subtler anomalies around 150 K and 50 K.
In SI,  we provide $M(T)$ for a second sample,  along with its derivative as a function of $T$ revealing clearer anomalies at $T_c = 204$ K (Fig. S1),  and also at $T = 156$ K and 78 K.  The sharpness and exact locations in temperature of both low-temperature anomalies are sample dependent, and contrast with the heat capacity data (Fig. S2) that reveals a single anomaly at $T_c = 215$ K  in good agreement with the change in slope seen in the resistivity $\rho_{xx}$ as a function of $T$ (Fig. 1c).  In our opinion, a plausible scenario for the anomalies in $M(T < T_c)$ are subsequent small variations in the tilt angle of the moments with respect to the out of the plane \emph{c}-axis.  Broad anomalies centered around these temperatures are also seen in the thermal transport, namely in the thermopower $S_{xx}(T)$ and thermal conductivity $\kappa_{xx}(T)$ measured on a Fe$_{2.84}$GeTe$_2$ single-crystal. This suggests changes in the scattering mechanisms (e.g. spin-spin scattering) affecting the charge carriers (Figs.  {\bf 1d-1e}). \textcolor{red}{But also, and perhaps most importantly, samples with markedly different Fe fractions, i.e. $(3.05 \pm 0.05)$ and $(2.84 \pm 0.05)$ show a couple of anomalies below $T_c$ implying that these are intrinsic to the material and not due to Fe deficiencies}.

We can further understand the behaviour of $M(T)$ via atomistic simulations\cite{Kartsev20,Wahab21,Augustin21} using the following spin Hamiltonian:
\begin{equation}
\mathcal{H} = -\frac{1}{2}\,  \sum_{i,j }  \mathbf{S}_i^\alpha \mathcal{J}_{ij}^{\alpha \beta} \mathbf{S}_j^\beta\:  -\frac{1}{2}\,  \sum_{i,j } K_{ij}\, (\mathbf{S}_i\, \cdot  \mathbf{S}_j\:)^2- \sum_{i} D_i (\mathbf{S}_i\, \cdot  \mathbf{e})^2 - \sum_{i} \mu_i  \mathbf{S}_i\, \cdot  \mathbf{B}
\label{gen_ham}
\end{equation}
where $i, j$ represent the atoms index,
$\alpha, \beta=x,y,z$, $\mathcal{J}_{ij}^{\alpha \beta}$
represents the exchange tensor that includes
the anti-symmetric exchange $-$
the Dzyaloshinskii-Moryia interaction (DMI),
$K_{ij}$ the biquadratic exchange interaction\cite{Kartsev20},
$D_i$ the uniaxial anisotropy,  which for Fe$_{3-x}$GeTe$_2$ is orientated out
of plane ($\mathbf{e}=(0,0,1)$) and
$\mathbf{B}$ the external magnetic field
applied during the field cooling (see additional details in SI).
It has been previously reported that higher-order
exchange interactions are fundamental in the description of
2D magnetic layers\cite{Kartsev20,Wahab21,Augustin21}.
We also noticed that Fe$_{3-x}$GeTe$_2$
develops substantial biquadratic interactions in its magnetic properties.
We estimated a magnitude of $K_{ij}$ within the range of 1.0 to 1.5 meV,
which follows those previously calculated for several
vdW materials\cite{Kartsev20}.  The inclusion of $K_{ij}$ provides
the best agreement between atomistic simulations and
measurements as can be observed in Fig. {\bf 1a}.
We barely found any difference between the calculated
$T_c=211.1$ K and that from measurements ($T_c=210$ K).
Indeed, we can further improve the calculation of $T_c$ through the
highly accurate magnetic susceptibility where no noticeable differences
are observable (see, Fig. S3 in SI).
In addition,  the curvature of $M(T)$ (Fig.  {\bf 1a})
is well represented using the model in
Eq.\ref{gen_ham} with $M(T) = [1-(T/T_c)]^\beta $,
where $\beta \approx 0.40$ from simulations and
(0.500$\pm0.003$) from neutron scattering measurements.
The slight smaller value of $\beta$ obtained from computations
suggests that additional effects in terms of the stabilisation of
the magnetic domains may play a role.  \\  \\
Figure 2a-2b show $M$ as a function of $\mu_0H$ applied along the
$c-$axis and $ab-$plane, respectively.  The data was collected
at several temperatures indicating that $i)$ the \emph{c}-axis is the
easy magnetization axis,  and $ii)$ the saturation moment
$\mu\approx$1.2 $\mu_B$ per Fe atom matches the
zero-field ordered moment extracted from neutron scattering
as $T \rightarrow 0$ K.  However the extracted
magnitude of $\mu$ is smaller than that ($\approx$3 $\mu_B$)
measured from the Curie-Weiss susceptibility and neutron scattering
data above $T_c$\cite{May1}.  We observed that for fields along the
$ab-$plane,  the integration of the neutron scattering
Bragg peak (whose intensity is given by both the lattice
and magnetic order scattering) reveals a continuous
decrease in the out-of-plane magnetic moment (Fig. 2c)
in favor of the planar one (Fig. 2d).  This is consistent
with the planar magnetization data as a function of $\mu_0H$
which increases continuously with field.
Our analysis suggests that the Fe moments
originally aligned along the $c-$axis cant progressively
towards the planar direction as the field increases.
This variation in spin orientation might contribute to generate non-coplanar
spin textures between the three Fe moments within
the formula unit (e.g.  spin chirality) leading to hitherto
unreported transport properties.

\subsection{Anomalous transport variables for magnetic fields
along the inter-planar direction and currents along a planar direction}

Usually the anomalous transport properties of magnetic compounds
are measured in a configuration implying thermal/electrical
currents oriented perpendicularly to the magnetic field which is
dictated by the texture of the Berry curvature at and below
the Fermi level\cite{AHE_review}.
For fields aligned along
the inter-planar direction\cite{Nernst},
the anomalous Hall resistivity $\rho_{xy}^A$ (Fig. 3a)
scales with the magnetization $M$ and is given
by $\rho_{xy} = \lambda M \rho_{xx}^n$ with $n \approx 2$,
where $\lambda$ represents the strength of the spin-orbit
coupling, $M$ the magnetization, and $\rho_{xx}$ the longitudinal resistivity.
For the many crystals measured in this work,
$\rho_{xy}^A(T)$ consistently tends to saturate
at a value within 10$-$12 $\mu \Omega$ cm.
We obtained a Hall angle $\theta_H = \sigma_{xy}^A/ \sigma_{xx} \simeq 0.04$ at $T = 25$ K
(Figs. 3a and 3b) roughly smaller than that previously reported ($\approx$0.07)\cite{Nernst}.
Nevertheless,  this value of $\theta_H$ is consistent with the Fe deficiency $\delta$,
which can cause Hall angles in the range of 0.04 to 0.085 \cite{Kim}.  Moreover,
$\rho_{xy}^A$ and $\sigma_{xy}^A$ decrease slightly upon
cooling below $T = 50$ K (Fig. 3b) consistently with previous works\cite{Nernst,Kim}.
The Nernst signal, where a transverse electric field $E_y$ is generated by a thermal gradient $\nabla T_x$ under an external field $\mu_0H$, is given by $S_{xy}^A = E_y^A/ (\mu_0H_z \nabla T_x)$, where $\mu_0H_z$ is the magnetic field along the inter-layer direction, was collected on the same crystal using the same electrodes (Fig. 3c) and reveals a maximum in the neighbourhood of 150 K followed by a change in sign below 50 K. At 150 K one obtains a Nernst angle $\theta_N = S_{xy}^A/ S_{xx} \simeq 0.073$ which is slightly smaller than 0.09 earlier reported\cite{Nernst}.

The Nernst signal results from a combination of terms in the thermoelectric $\alpha_{\alpha\beta}$
and charge conductivity $\sigma_{\alpha\beta}$ ($\alpha,\beta=x,y,z$) tensors:
\begin{equation}\label{Sxy}
S_{xy} = \frac{\alpha_{xy}\sigma_{xx} - \alpha_{xx}\sigma_{xy}}{\sigma_{xx}^2 + \sigma_{xy}^2}
\end{equation}
where the transverse thermoelectric conductivity is given by:

\begin{equation} \label{alphaxy}
\alpha_{xy} = \frac{e k_B}{\hbar}\sum_n \int_{\text{FS}} \frac{d^3k}{(2 \pi)^3}\Omega^z(\textbf{k})s(\textbf{k})
\end{equation}
and $\Omega^z$ is the $z$ component of the momentum
integrated Berry curvature,
and $s(\textbf{k}) = -f(\textbf{k})\ln f(\textbf{k})-(1-f(\textbf{k}))\ln(1-f(\textbf{k}))$
is the entropy density with $f(\textbf{k})$ as the Fermi distribution function.
The integral is computed over the Fermi surface (FS),
and the summation performed for every occupied band $n$.
Given that $\alpha_{xx}$ is negative over the entire $T$ range (Fig. 1e)
and that in contrast $\rho_{xy}$ is positive or dominated by holes\cite{WangY} (Figs. 3a and 3b),
the only term that can lead to a change in the sign of $S_{xy}$ in Eq. \ref{Sxy}
is the transverse thermoelectric conductivity term $\alpha_{xy}^A$.
To expose this point,  we use $S_{xy}^A(T)$ (Fig. 3d)
and Eq. \ref{Sxy} to calculate $\alpha_{xy}^A(T)$ via:
$\sigma_{xx}(T) = \rho_{xx}^{-1}(T)$ (Fig. 1c),
$\alpha_{xx}(T)$ (Fig. 1e),  and $\sigma_{xy}^A(T)$.
It results that $\alpha_{xy}^A(T)$ (Fig.  3d) follows
the overall behavior of $S_{xy}^A(T)$ also changing its sign below $T\approx$30 K.
Equation \ref{alphaxy} implies that this change in
the sign of $\alpha_{xy}^A(T)$ ought to result
from a sharp reconfiguration of the Berry
curvature at the Fermi level occurring below $\approx$50 K.
A topological transition either associated with a new magnetic texture
or an electronic Lifshitz transition likely coupled to the magnetic order is feasible
to occur.  In contrast,  the change in the sign of $\alpha_{xy}$ observed at $T_c$
is associated to a change in the entropy density.

The final anomalous transport variable,  that is,
the thermal Hall or Righi-Leduc conductivity
$\kappa_{xy}^A = j_{Qx} / \nabla T_y$ (where $j_{Qx}$ is
the heat current) also reveals broad anomalies
 as a function of $T$ (Figs. 3e-3f).
 $\kappa_{xy}^A(T)$ increases sharply below $T_c$
and display maximum values around
$\approx$150 and $\approx$55 K.
This behaviour suggests a possible role
of magnons in the compound thus supporting
the notion of pronounced changes in the
scattering mechanisms at these temperatures.
 The maximum value of the Rhighi-Leduc coefficient
 at $\approx$150 K ($\kappa_{xy}^A \simeq 48 \times 10^{-3}$ W/mK)
 is larger than that at Mn$_3$Sn at $T=300$ K\cite{Behnia2}.
 Defining a thermal Hall angle as $\theta_{\text{TH}} = \kappa_{xy}^A/\kappa_{xx}$ one obtains $\theta_{\text{TH}} = 0.016$ around 150 K.
 Overall, the values measured by us for these
 anomalous transport variables are comparable to
 those reported for other topological compounds
 such as Mn$_3$Sn \cite{Satoru, Ikhlas, Behnia2}
 and Mn$_3$Ge \cite{Behnia}.

\subsection{Wiedemann-Franz law}
Following the anomalies observed in $\kappa_{xy}$,
it is pertinent to ask if the observed anomalous transport
quantities in Fe$_{3-x}$GeTe$_2$ would satisfy the Wiedemann-Franz (WF) law,
i.e. $\kappa_{xy}/T \sigma_{xy} \simeq L_0$ where $L_0$
is the Lorentz number.  This would imply that
the same carriers would transport heat and charge.
Such issue was recently addressed for
both Mn$_3$Sn\cite{Behnia2} and
Mn$_3$Ge\cite{Behnia} compounds that are claimed
 to display Weyl nodes relatively close
 to their Fermi level \cite{Kuroda}.
 The former is claimed to satisfy the WF-law
 whereas the latter shows a pronounced deviation
 approaching room temperature
 that is not attributed to inelastic scattering\cite{Behnia}.

The aforementioned anomalies that are
quite marked in $\kappa_{xy}/T$ (Fig. 4a)
contrast with the smooth evolution of
$\sigma_{xy}^{A} = - \rho_{xy}^{A}/(\rho_{xy}^2+ \rho_{xx}^2)$ (Fig. 4b)
upon decreasing $T$.
Although both quantities were measured on the same crystal,
which precludes the errors inherent to the measurement of its
geometrical factors,  there is an intrinsic experimental error within
these measurements as illustrated at $T = 250$ K (Fig.  4c).
Within the conservative error bars estimated by us,  the anomalous
quantities in Fe$_{3-x}$GeTe$_2$ seem to satisfy the WF-law.

\subsection{Anomalous transport variables for magnetic fields along the gradient of the chemical potential}
Subsequently,  we discuss the most intriguing aspects of
the anomalous transport properties of Fe$_{3-x}$GeTe$_2$.
When the magnetic field is oriented along the gradient of
chemical potential (parallel to either the applied current or thermal gradient),
we observe anomalous planar Hall and anomalous planar heat transport variables
that are truly antisymmetric as a function of field orientation (see Fig. 5).
This observation is not to be confused with the conventional planar Hall-effect discussed for example, in the context of Weyl semimetals \cite{Burkov}, that is a measure of the anisotropy of the magnetoresistivity as a function of field orientation and therefore is an even signal of the magnetic field.  In contrast, our observations are an odd function of magnetic field.

At lower $T$s the maximum value of $\rho_{xy}^A$, that ranges from $\approx 3$ $\mu \Omega$ cm and $\approx 8$ $\mu \Omega$ cm is sample dependent, indicating that sample quality, the exact slight deficiency in the occupancy of Fe, and errors in the precise determination of $t$ play a role on the extracted numbers. We performed detailed checks for experimental artifacts, more specifically whether the antisymmetric signal could be caused by slight misalignments of the magnetic field (see, Fig. 5, S4 and S5 in SI), but in the process we became aware of Ref. \cite{PlanarTHE} showing similar experimental results. Ref. \cite{PlanarTHE} suggests that this unexpected observation would result from an internal gauge field resulting from a complex spin texture associated either with the formation of skyrmions \cite{Wu,Ding,skyrmion3} or a complex non-coplanar spin texture \cite{Fe312_DMI}. We observed this anomalous planar Hall-effect in more than 6 samples of different geometries and thicknesses (see, Figs.5, S4 and S5), despite differences in sample quality and Fe deficiency, suggesting that this effect indeed has a topological origin. Concerning this anomalous Hall signal for currents along the magnetic-field (Figs. 5, S4 and S5), one of our initial concerns was inhomogenous current distribution with a fraction of it flowing along the inter-planar direction. Although this might create a Hall-like signal for currents originally expected to flow along a planar direction aligned along the field, it would not explain the observation of this anomalous Hall signal for currents flowing along a planar direction but oriented perpendicularly to the field.

We also measured the anomalous variables of Fe$_{3-x}$GeTe$_2$ for fields applied along the gradient of the chemical potential through pulsed methods. For these measurements, we used an electrical current (or thermal heat) pulse method to minimize sample self-heating within the vacuum cell. Although $\rho_{xy}^A$ for $j \| \mu_0H$ collected through this experimental method is subjected to a poorer signal to noise ratio (Fig. 6a), it yields essentially the previously discussed behavior (Fig. 5). For instance, one sees that $\rho_{xy}^A$ at the lowest $T$s displays a maximum in the neighborhood of $\mu_0 H \approx 4 $ T (Fig. 6b). This behavior is followed by the Nernst signal $S_{xy}^A(\mu_0H \| j_Q)$ collected on the same sample (Fig. 6c) which also shows a maximum around the same field value. Although not seen in this data-set, due to limitations in the signal to noise ratio, we also observed a change in the sign of $S_{xy}^A$ upon cooling below $T = 50$ K in another crystal using the constant heat gradient method and a different experimental set-up (Fig. S6). Finally, $\kappa_{xy}^A(\mu_0H \| j_Q)$ mimics  $S_{xy}^A(\mu_0H \| j_Q)$, with the maxima seen in both thermal transport variables as a function of $\mu_0H$ disappearing as $T$ is lowered. This contrasts with $\rho_{xy}^A (\mu_0H \| j)$ whose maxima increase as $T$ is lowered down to $\approx 50$ K, from which point $\rho_{xy}^A (\mu_0H \| j)$ decreases slightly.

We understand the shape of the anomalous planar variables (Figs. 5 and 6) as resulting from a magnetic-field induced canted and non-coplanar spin texture that leads to a field-dependent spin chirality $\chi_{ijk}$ that acts as an effective gauge field and reaches a maximum value in the neighborhood of $\mu_0H \simeq 4.5$ T. Remarkably, as the moments align along the field thus suppressing $\chi_{ijk}(\mu_0H)$, $\rho_{xy}^A$ remains finite instead of reaching zero. We speculate that this results from the intrinsic topological nature of the electronic band structure of Fe$_{3-x}$GeTe$_2$. To support this assertion, we point to the observation of an anomalous planar Hall signal already in the paramagnetic state, or for 220 K $< T \leq $ 300 K (Figs. 7, S4 and S5). In contrast to what is seen in the ferromagnetic state, where $\rho_{xy}^A$ shows a maximum somewhere between $\mu_0H = 4$ T and 5 T (or where the planar magnetization begins to saturate), this Hall like signal initially increases linearly with field but tends to saturate as $\mu_0H$ increases beyond 6 T. This implies, within the gauge field scenario proposed by Ref.\cite{PlanarTHE}, that Fe$_{3-x}$GeTe$_2$ would already display a topological non-trivial character in its paramagnetic state. Given the absence of magnetic order, its origin would have to rely on its electronic band structure due, for example, to the existence of a Dirac nodal line as proposed by Ref. \cite{Kim}, although the electronic structure calculations in Ref. \cite{Kim} explicitly take into account the FM order. Therefore, Fe$_{3-x}$GeTe$_2$ might provide an unique example of a compound displaying coexisting mechanisms affecting the texture of its Berry phase for fields applied along a planar direction.

\begin{figure}[htb]
	\begin{center}
		\includegraphics[width = 6.5 cm]{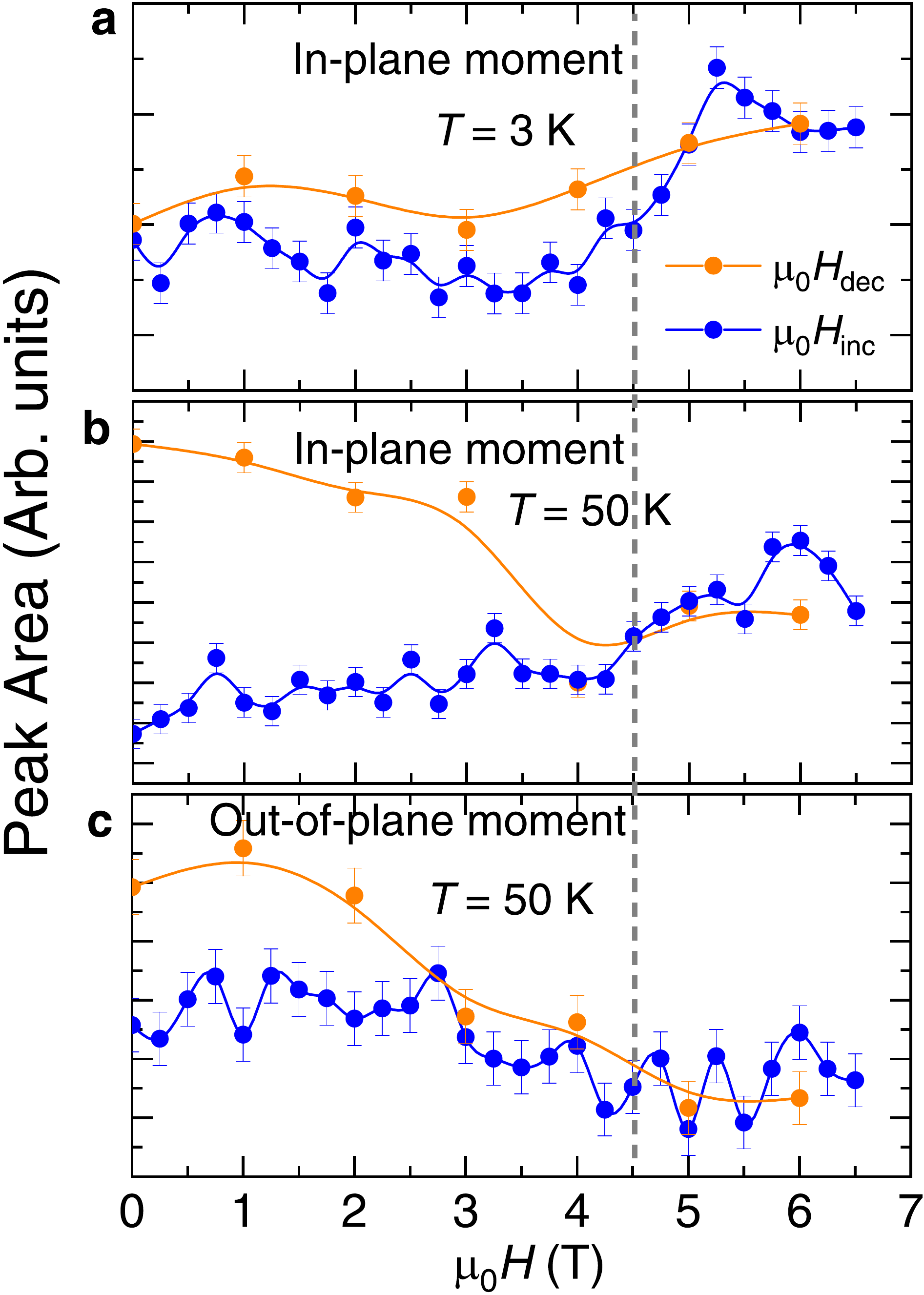}
		\caption{\textbf{Evidence for a change in the magneto-crystalline anisotropy at $T \simeq 50$ K.}  Integrated area of the Bragg peaks for both increasing (blue) and decreasing (orange) magnetic fields for {\bf a,} the (002) peak at $T = 3$ K, {\bf b,} the (002) peak at $T = 50$ K,  and {\bf c,} the (100) peak at $T = 50$ K.
Increasing and decreasing field sweeps were taken at subsequently increasing temperatures, alternating between the (100) and (002) peaks at each field point.}
	\end{center}
\end{figure}

\subsection{Non-trivial topological spin textures}

\begin{figure}[!h]\centering
 \includegraphics[width=1\columnwidth]{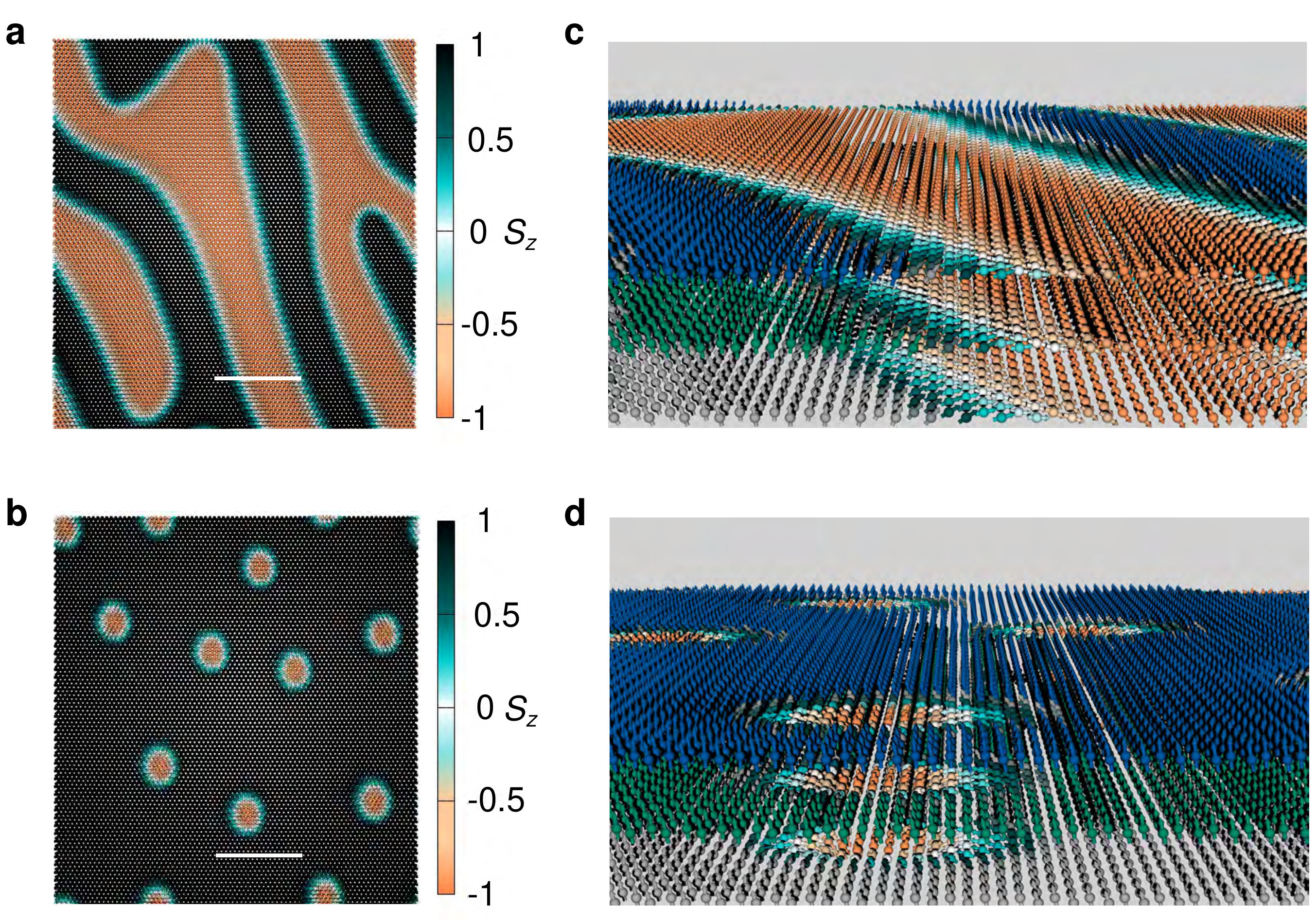}
 \caption{{\bf Stripes and skyrmions on Fe$_3$GeTe$_2$.}
{\bf a-b,} Snapshots of the magnetic ground-state at 0 K
obtained during the cooling process at zero and a finite field of 2 T,  respectively.
The color scheme on the right shows the variation of the out-of-plane magnetisation $S_z$.
The in-plane sizes are 32 nm$\times$ 34.55 nm along $x$ and $y$.  The white bars correspond to a length scale of 7.0 nm.
{\bf c-d,} Side views of {\bf a} and {\bf b} at the location of the white bars.
The color of the saturated region,  where the magnetisation is larger
than $S_z=0.8$ has been altered to gray, green and blue to emphasize the different Fe layers.
Here, $S_z=1.0$ implies full polarization of the Fe moments along the field.
With an applied field of 2 T ({\bf b} and {\bf d}), N\'eel skyrmions are observed.
The topological number $Q$ of these magnetic structures is calculated as $Q=0.05$ ({\bf a,c}) and  $Q=-10.63$ ({\bf b,d}).
\label{theory1}
}
\end{figure}

We can shed some light on the unique magnetic
features observed on the experimental results
using atomistic spin dynamics\cite{Kartsev20,Wahab21,Augustin21}
considering the spin Hamiltonian in Eq.\ref{gen_ham}.
The system is thermally equilibrated above the Curie temperature ($T=300$ K)
and then linearly cooled for 2 ns to a temperature of
$T=0$ K,  followed by a 1 ns relaxation at zero-temperature (Figure \ref{theory1}).
Following the results of $M(T)$ and the importance of
higher-order interactions (Fig.  1{\bf a}) in the magnetic properties of Fe$_3$GeTe$_2$,
a biquadratic nearest-neighbours exchange of $K_{ij}=1$ meV
has been used.  We also included an in-plane DMI value of 10$\%$
of the Heisenberg exchange implemented following the symmetry
presented in Laref \emph{et al}. \cite{Laref}.
In the absence of an applied field,
the ground-state is a stripe domain phase (Fig.  \ref{theory1}{\bf a,c}) with a
topological number $Q$\cite{rozsa_prb} (see section \ref{skyrmionN} in {\it Methods} for details)
that indicates a trivial topology ($Q=0.05$) or no topological protection.
Once a magnetic field is applied perpendicularly to the
surface  N\'eel skyrmions are observed
(Fig.  \ref{theory1}{\bf b,d}) resulting in a
topological number $Q=-10.63$ in agreement with the amount
observed on the top view (Fig. \ref{theory1}{\bf b}).
We also noticed that at each unit cell of monolayer Fe$_3$GeTe$_2$ composing the bulk,
there is the formation of three skyrmions resulting of the two inequivalent Fe
atoms.  That is,  one skyrmion per Fe atomic layer
disposed parallel to each other in the crystal structure
(Fig. \ref{theory1}{\bf d}).  In such sandwich arrangement,
substantial interactions occur between the skyrmions
along the interlayer direction which aligned them roughly at the
same position but at different heights.
The skyrmion diameter $D_{sk}$ can be extracted numerically from
the computations (see section \ref{skyrmionR} in {\it Methods} for details)
resulting in an average value of $D_{sk}=2.424$ nm.
This skyrmion size can be directly related to the amount of
DMI included in our simulations
which can be additionally tuned
at different magnitudes\cite{Wang:2018wq}.
We however used a more qualitative approach
to show the existence of non-trivial spin textures into the system.
\begin{figure}[!h]\centering
 \includegraphics[width=1\columnwidth]{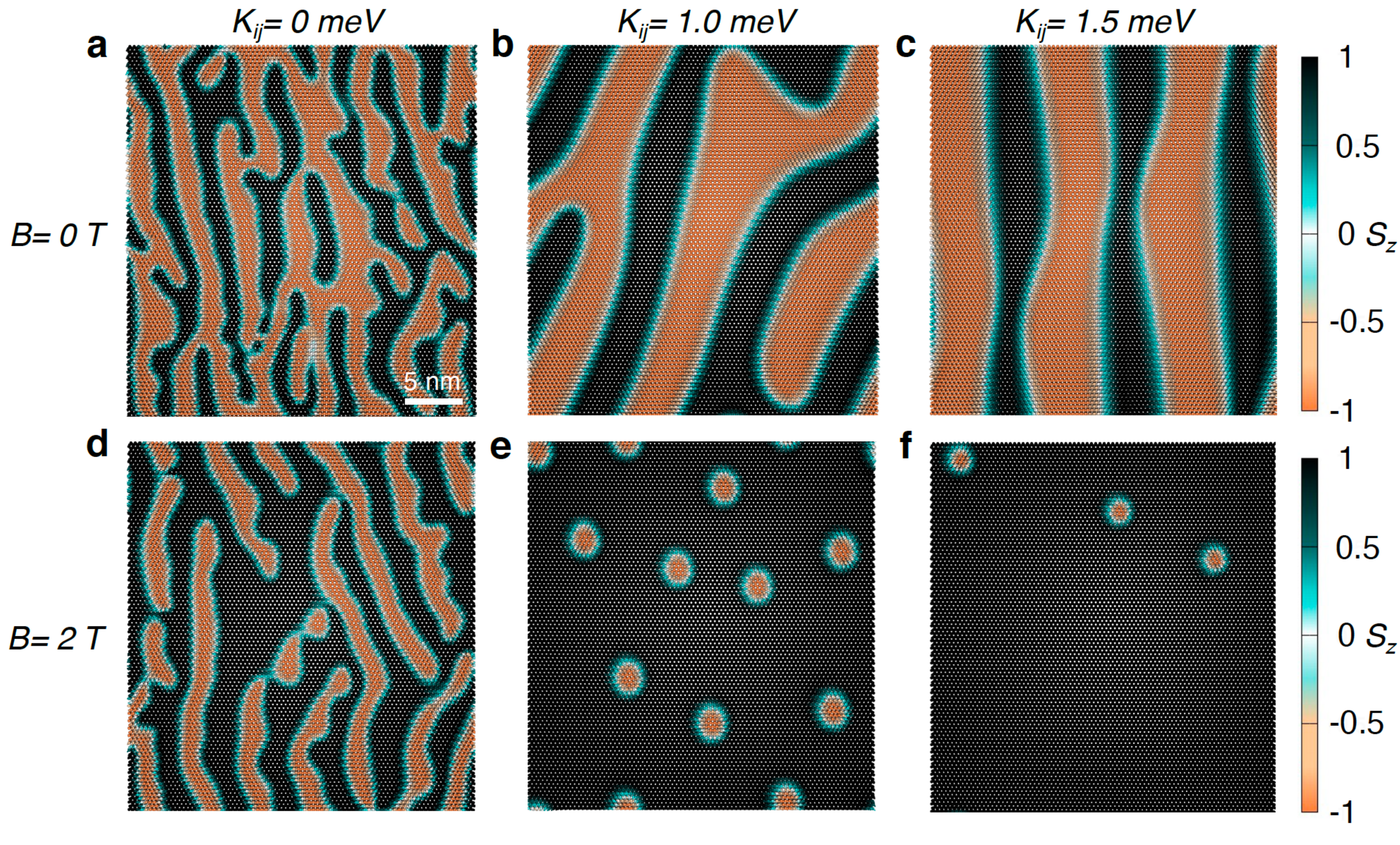}
 \caption{{\bf Biquadratic exchange driven skyrmion formation on Fe$_3$GeTe$_2$. }
{\bf a-c,} Snapshots of the spin dynamics at 0 K and zero magnetic field ($B=0$)
of bulk Fe$_3$GeTe$_2$ taking into account
different amount of biquadratic exchange $K_{ij}$
as 0 meV, 1.0 meV and 1.5 meV,  respectively.
The out-of-plane component of the magnetisation $S_z$ (color scale)
is used to track down the spin distribution.
{\bf d-f,} Similar as {\bf a-c,} but at an external field of $B=2$ T.  In the absence of
biquadratic exchange ($K_{ij}=0$ meV) no skyrmion states are observed
with or without an applied field ({\bf a,d}).
At finite $K_{ij}$ the groundstate is drastically
changed with the appearance of skyrmions under $B$ ({\bf e,f}).
By increasing $K_{ij}$,  the number and size of the skyrmions diminish whereas
in the absence of an applied field,  an increment
in the characteristic length-scale of the
stripe domains is observed ({\bf b ,c}).
The correspoding topological number $Q$ is: {\bf a,} $Q=2.191$,  {\bf b,} $Q=0.05$,  {\bf c,} $Q=0.0008$,
{\bf d,} $Q=-6.929$,  {\bf e,} $Q=-10.63$ and {\bf f,} $Q=-2.974$.  \label{theory2}
}
\end{figure}

It is worth mentioning that the
formation of skyrmions on Fe$_{3-x}$GeTe$_2$
not only depends on DMI,  but also on the
biquadratic exchange $K_{ij}$ (Eq. \ref{gen_ham})
as shown in Figure \ref{theory2}.  As long as $K_{ij}=0$ no skymions
are observed under zero (Fig. \ref{theory2}{a}) or
finite magnitudes of $B$ (Fig. \ref{theory2}{d}).
The scenario changes substantially as a finite
value of $K_{ij}$ is included into the system with the
stabilisation of skyrmions throughout the layer once
the magnetic field is switched on (Fig. \ref{theory2}{e-f}).
The different values of $K_{ij}$ induce variations
on the size of the stripe domains as well as on the skyrmions
and their topological number.  At $B=0$ T,  the
lengthscale of the stripe domains increases
with the increment of $K_{ij}$ (Fig. \ref{theory2}{\bf b-c})
become broader and more extended over the entire crystal.
At $B=2$ T,  the skyrmions decrease in diameter and number which suggest a
critical interplay between magnetic field and higher-order exchange processes.
On that,  the biquadratic exchange favors a perpendicular alignment
of the spins in order to minimise the total energy which
interacts with the in-plane DMI allowing
the spins to precess around some perimeter.
The higher the value of $K_{ij}$ the smaller the precession of
the spins as they will be pointing along the same direction (e.g.  out-of-plane).
This implicates shorter radius and consequently less possibility of
skyrmion nucleation.

\section{Concluding remarks}
The Fe$_{3-x}$GeTe$_2$ transport data discussed here can only be understood in terms of a gauge field intrinsic to the electronic band structure of Fe$_{3-x}$GeTe$_2$ that acts as an effective ``band magnetic field" that bends the electronic orbits to yield, for example, an anomalous Hall-like signal in the paramagnetic state in the absence of interaction between the electrical current and the external magnetic field. At lower temperatures within the ferromagnetic state, this anomalous and antisymmetric planar Hall signal is accompanied by anomalous planar Nernst and Righi-Leduc effects, which to the best of our knowledge have yet to be reported. We propose that the local Dzyaloshinskii-Moriya interaction, claimed to stabilize skyrmions \cite{Ding, Pei, skyrmion3} and chiral spin spirals at its surface \cite{Fe312_DMI}, favors a magnetic-field dependent canted and non-coplanar spin structure that scatters the charge carriers affecting their Berry phase and leading to the novel effects observed by us. To support this point, we used neutron diffraction to measure magnetic order with fields along the planar direction (Fig. 8). For fields along the planar direction, the area of the diffracted neutron peak corresponding to the in-plane moment increases sharply in the neighborhood of $\mu_0 H \approx 4$ K (Fig. 8a) which is consistent with a metamagnetic transition leading to the pronounced dips observed in the anomalous planar transport quantities (Figs. 5 and 6). At $T = 50$ K, and for fields along the \emph{a}-axis, neutrons reveal a very pronounced hysteresis between field increasing and decreasing sweeps (Figs. 8b and 8c), pointing to either a temperature-dependent change in the magnetic anisotropy, a metamagnetic transition, or the emergence of another magnetic order. The elucidation of its origin will require a small angle neutron scattering study. However, these observations, combined with a temperature dependent magnetic anisotropy found in our neutron study (Fig. S7), suggest changes in the magnetic domain structures and their textures, possibly leading to local regions with chiral spin order \cite{Fe312_DMI, Elbio} instead of a simple collinear one, or even the possibility of a transition between both states as function of $T$. For example, as discussed in Ref. \cite{Elbio}, conical spin spirals would naturally lead to a planar topological Hall-effect (due to spin-chirality), as exposed here. \textcolor{red}{We re-emphasize that heat capacity measurements do not support the notion of thermodynamic phase-transitions, as one would expect for electronic, magnetic, or structural transitions, albeit these could be subtle requiring further investigations.}

Regardless of the precise field-induced spin texture, we have found that it also leads to planar topological Nernst and thermal Hall-effects that have yet to be reported or discussed in the literature. Our Monte Carlo simulations including Dzyaloshinskii-Moriya and biquadratic exchange interactions correctly capture the formation of the labyrinthine domains previously observed under zero magnetic field as well as the development of N\`{e}el skyrmions upon field application \cite{Ding,Pei,skyrmion3}. Most importantly, upon application magnetic fields along a planar direction, the biquadratic exchange interaction favors the stabilization of the labyrinthine domains characterized precisely by a chiral spin texture (see, Fig. S10) associated with Bloch domain walls, thus providing the necessary ingredient, i.e. spin chirality, for the topological transport variables reported here. \textcolor{red}{In fact, during the review process, we became aware of Ref. \onlinecite{Bloch_domains} that confirms, via Lorentz microscopy, the spin textures found in our simulations for fields along a planar direction.} Among the Supplementary Materials, we also include three movies, or Monte Carlo simulations, illustrating the formation of these domains upon decreasing temperature, and the creation of skyrmions upon increasing the external field applied along the \emph{c}-axis, respectively. Therefore, this system is prone to the development complex spin textures characterized by a finite spin-chirality.

Notice that for tilted magnetic fields one tends to observe the development of a pronounced peak in $\rho_{xy}^A$ (Fig. S8) requiring fields in excess of $\mu_0 H = 10$ T for its suppression. This peak is likely to correlate with the development of the aforementioned complex textures, e.g. skyrmions, skyrmion tubes, and spin spirals \cite{Elbio}, as implied by our Monte Carlo simulations, but become suppressed at higher magnetic fields where the Hall effect becomes dominated by the anomalous contribution. Additional neutron studies will be required to elucidate this behavior. However, we propose that deviations from spin collinearity as seen in the surface of this compound \cite{Fe312_DMI}, are likely to be responsible for the seemingly topological phase transition seen by the Nernst signal in the neighborhood of $T = 50$ K. Anomalies associated with this transition are seen in all other transport variables with the exception of the resistivity. These subtle transitions are not detected in the heat capacity either.

Finally, the Nernst effect is often used as a probe for topological excitations in quantum materials providing a means to convert heat into electricity \cite{Kuroda,Akito}. Our study reveals a new way to detect, or expose such excitations as well as a new geometry for heat conversion that expands the horizon of thermoelectric technology.
Notice, that the Curie temperature of Fe$_{3-x}$GeTe$_2$ was shown to increase well-above room $T$ when grown on Bi$_2$Te$_3$ \cite{RoomT} thus offering the possibility of studying the interplay between magnetic and electronic topology under ambient conditions. Such heterostructures could also open a new avenue for efficient heat conversion at, or above, room temperature.

\setlength{\parindent}{0pt}
\section*{Materials and Methods}

\setlength{\parindent}{0pt}
\subsection*{Crystal synthesis} Nominal Fe$_{3-x}$GeTe$_2$ crystals were grown either through a chemical vapor transport (CVT) method using iodine as
the transport agent or via Te flux method. For CVT grown crystals, the excess iodine was removed prior characterization through a bath and rinse cycle
in acetone and isopropanol. \textcolor{red}{The composition of the measured crystals was determined through energy-dispersive x-ray spectroscopy, yielding values for the Fe fraction ranging from $x_{\text{Fe}} \approx (2.80 \pm 0.05)$ to $\approx (3.05 \pm 0.05)$, with $\Delta x_{\text{Fe}} = 0.05$ being the typical standard deviation in any given crystal.}

\setlength{\parindent}{0pt}
\subsection*{X-ray diffraction} Single-crystal x-ray diffraction was performed on a Rigaku-Oxford Diffraction Synergy-S diffractometer equipped with a HyPix detector and a monochromated Mo-$K_{\alpha}$ radiation source ($\lambda$ = 0.71073 \AA). Single crystals were mounted on a nylon loop with help of Parabar oil (Hampton Research). The data were collected as $\omega$-scans at $0.5^{\circ}$ step width, and the unit cell was refined with the CrysAlis software package. X-ray diffraction confirmed that the crystal structure adopts the space group space group 194 ($P6_3/mmc$) with lattice parameters $a =3.9747(12)$ \AA, $c=16.370(5)$ \AA $\text{ }$ and angles $\alpha = 90.01(2)^{\circ}$, $\beta = 90.03(2)^{\circ}$, and $\gamma = 119.98(2)^{\circ}$, where the numbers in parenthesis indicate the uncertainty in the last digits.

\subsection*{Neutron diffraction measurements} Neutron diffraction measurements were performed on the BT-4 triple axis spectrometer at the NIST Center for Neutron Research. The neutron energy was fixed at 14.7 meV ($\lambda$ = 2.359 \AA), with pyrolytic graphite filters placed between the sample and analyzer to remove higher-order contamination. The sample was sealed in a helium environment and placed in a closed-cycle refrigerator to vary the temperature between 4 and 300 K. Temperature-dependent measurements were performed without a magnetic field; field-dependent measurements were performed with a magnetic field between 0 and 7 T applied within the \emph{ab}-plane, approximately perpendicular to the (100)-axis.

\subsection*{Thermal transport measurements} Thermal conductivity and the thermal Hall-effect were measured using a one-heater three-thermometer method. Additional electrical contacts allowed us to measure four-probe resistivity, Hall-effect, Seebeck and Nernst effects simultaneously. For the thermal transport measurements a heat pulse was applied in order to generate a longitudinal thermal gradient corresponding to a $\approx 3$ \% of the sample base temperature. After applying the heat pulse, the temperature of all three thermometers were monitored until they reached a stable condition (defined as a rate of less than 1 $\mu$K/s) averaged over 15 seconds.
Typical timescales were 5 s to 10 s for temperature rise and 30 s to 60 s for its stabilization. A step-wise increase in heat was also applied to generate corresponding step-wise thermal gradients, from which a linear relation between the measured values (e.g. thermal electromotive force as a function of temperature gradients for Seebeck and Nernst effects; temperature gradients as a function of heat power) were used to obtain the relevant thermal transport variables. The results from the both methods are practically identical. The measurements were performed in Quantum Design Physical Properties Measurement system (Quantum Design-PPMS), which allowed in-situ calibration of thermometers in the presence of exchange gas followed by the thermal measurements under high vacuum.

\setlength{\parindent}{0pt}
\subsection*{Magnetotransport measurements} Conventional magnetotransport experiment were performed in a physical property measurement system (Quantum Design - PPMS) under magnetic fields up to $\mathrm{\mu_0} H = 9$ T and temperatures as low as 2 K using a custom made probe and external electronics. A $^3$He cryostat where the samples were immersed in liquid $^3$He, in combination with a rotating probe was used for high field experiments up to $\mu_0H = 31$ T at temperatures down to 0.35 K.

\setlength{\parindent}{0pt}
\subsection*{Magnetization and heat capacity measurements}
Magnetization measurements under fields up to $\mathrm{\mu_0} H = 7$ T were performed in a commercial superconducting
quantum interference device magnetometer (Quantum Design - SQUID). Heat capacity measurements through the thermal relaxation method were collected in a Quantum Design-PPMS.

\setlength{\parindent}{0pt}
\subsection*{Atomistic spin dynamics}
The magnetization as a function of temperature were calculated
by using Monte Carlo method as described in Ref. \cite{Kartsev20, Wahab21,Augustin21}.
A 288$\times$166$\times$1 model super-cell (191232 sites in total)
based on the rectangular unit cell of hexagonal lattice with
periodic boundary conditions was used.
The spin dynamics calculations are performed using atomistic spin dynamics
within the stochastic Landau-Lifshitz-Gilbert equation\cite{Ellis} integrated
with the Heun numerical scheme and 1 fs timestep \cite{Wahab21}.
Equilibrium temperature dependent properties are computed using
the Monte Carlo Metropolis method integrated with the adaptive move
algorithm\cite{Alzate_Cardona_2019}.
Tables \ref{simulation parameters}-\ref{summary_compare_models}
summarise the parameters used in the simulations:
\begin{table}[!h]
\centering
\fontsize{9}{8}
\begin{tabular}{c|l|l|l}
Quantity~ & Symbol~ &quantity~ & units~ \\
 \\\hline
Timestep & $ts$ &0.5 &fs $(10^{-15} s)$\\
Damping & $\alpha$ &0.1  &\\
Magnetic moment Fe$_{I,II}$ & $\mu$ & 1.95, 1.56 \cite{verchenko2015ferromagnetic} & $\mu_B$\\
Uniaxial anisotropy            & $D_i$ &5.345  $\times 10^{-23}$ \cite{Deng} &J/link \\
Unit cell constants           & $x,y,z$ &7.07, 12.26, 25 &A \\
Maximum temperature           & $T_{max}$ & 300&K \\
Minimum temperature           & $T_{min}$ & 0&K \\
Applied field           & $B$ & 0.0,  ~2.0&T \\
\end{tabular}
\caption{Simulation parameters \textcolor{red}{used in the atomistic calculations to capture the spin textures in Fe$_{3-x}$GeTe$_2$.}}
\label{simulation parameters}
\end{table}
\begin{table}[!h]
\centering
\fontsize{9}{8}
\begin{tabular}{c|l|l|l}
Interaction~ & Fe sub-lattice~ & J (meV/link)~ &DMI (meV/link)\\
 \\\hline
J$_1$   & I-I &6.48 &0.0\\
J$_2$   & I-II &27.9 &2.79\\
J$_3$   & I-I &-5.79 & -0.579\\
J$_4$   & II-II &-4.64 & -0.464\\
J$_5$   & I-I &1.23 & 0.0\\
J$_6$   & I-II &-0.12 & 0.0\\
J$_z1$   & I-I &-1.93 & 0.0\\
J$_z2$   & I-I &0.21 & 0.0\\
J$_z3$   & I-II &1.44 & 0.0\\
\end{tabular}
\caption{Exchange parameters at \textcolor{red}{different number of nearest neighbors $n$ along the in-plane ($J_n$ from $n = 1$ to 6) and out-of-plane ($J_{zn}$ from $n = 1$ to 3) directions} used for the Fe$_3$GeTe$_2$ simulations, as extracted from Deng \emph{et al}. \cite{Deng}.
The DMI value has been set to $10\%$ of the exchange, with the symmetry as given in Ref. \cite{Laref}.}
\label{summary_compare_models}
\end{table}
\setlength{\parindent}{0pt}
\subsection*{Topological number calculation}\label{skyrmionN}
To calculate the skyrmion or topological number we follow the procedure
shown in Ref. \cite{rozsa_prb}.  The hexagonal lattice is split
into nearest-neighbour triangles of spins.
The topological charge associated with the system region
defined by spins $S_1-S_4$ will be given by the summation
between two neighbour triangles ($\omega_1^i, \omega_2^i$)
as shown in Fig. S9 in the SI,  where the spherical
area of each triangle is \cite{eriksson_1990}:
\begin{equation}
    \omega=2 \arctan \left( \frac{S_1 \cdot (S_2 \times S_3)}{1+S_2 \cdot S_3 + S_3 \cdot S_1 + S_1 \cdot S_2}\right)
\end{equation}
The total topological number will be given by the summation over all triangles in the system:
\begin{equation}
   Q= \frac{1}{4 \pi} \sum_i(\omega_1^i + \omega_2^i)
\end{equation}
The negative sign of the topological number is given by the orientation
of the core of the spin structures.  The topological number has
been calculated per each individual atomic layer,
then the average between the three atomic layers in the Fe$_{3-x}$GeTe$_2$ system has been performed.


\setlength{\parindent}{0pt}
\subsection*{Skyrmion radius calculation}\label{skyrmionR}

In order to calculate the skyrmion radius,
the profile of the magnetisation has been fitted to\cite{Wang:2018wq,Braun1994}:
\begin{equation}
    S_z(x)=\cos \left[2 \arctan \left({\sinh(R/w)}/{\sinh((x-x_0)/w)}\right)\right]
     \label{eq_skyrmion_radius}
\end{equation}
where $x_0$ is the position of skyrmion,  $R$ its radius and $w$ the width
of the domain wall.  The radial direction the magnetisation profile provides a 360$^{\rm o}$
N\'eel domain wall \cite{Wang:2018wq,Braun1994}.  Once the temperature is cooled in Fe$_{3}$GeTe$_2$,
we find multiple skyrmions throughout the layers.
In order to calculate numerically the average radius of the skyrmions,
we determine the region along the x-axis where $S_z$ is minimum
and fit the profile for $y=constant$ via Eq. \ref{eq_skyrmion_radius}.
The average skyrmion diameter $D_{sk}=2R$ obtained for the first
layer of skyrmions that consists of 11 skyrmions shown in panel c),
Fig. \ref{theory1} gave a value of $D_{sk}=2.424$ nm.

\setlength{\parindent}{0pt}
\section*{Supplementary Material}
See the Supplementary Material for single-crystal x-ray refinement parameters from a Fe$_{3-x}$GeTe$_2$ sample.
magnetization $M$ as function of the temperature $T$ for another crystal, heat capacity as a function of the temperature, calculated magnetic susceptibility $\chi$ based on the model described in the main text, angular dependence of the anomalous planar Hall-effect at different temperatures, amplitude of the anomalous planar Hall effect as function of the temperature and for several values of the external field, anomalous Nernst $S_{xy}^A$ signal as a function of $\mu_0 H$ for another Fe$_{3-x}$GeTe$_2$ crystal collected under the constant heat gradient method and for heat currents $j_Q \| \mu_0 H$, magnetic anisotropy as a function of the temperature according to neutron scattering, anomalous Hall resistivity $\rho_{xy}^A$ as a function of field $\mu_0 H$ collected at $T = 410$ mK and for several angles $\theta$ between $\mu_0 H$ and the interlayer \emph{c}-axis, schematics illustrating how the skyrmion number was calculated through the Monte Carlo simulations, and Monte Carlo snapshots indicating how spin spirals form in Fe$_3$GeTe$_2$ upon application of a planar magnetic field.

\setlength{\parindent}{0pt}
\textbf{Correspondence and requests for materials} should be addressed to L.B.

\setlength{\parindent}{0pt}
\section*{Acknowledgments} We acknowledge fruitful discussions with Prof. Kun Yang and Prof. Satoru Nakatsuji and Prof. Leon Balents. L.B. is supported by DOE-BES through award DE-SC0002613. This research was supported in part by the National Science Foundation (award DMR-1905499 to M.S.). J.Y.C. acknowledges support from NSF-DMR-1700030. D.R. acknowledges support from the NHMFL Visiting Scientist Program. The National High Magnetic Field Laboratory acknowledges support from the US-NSF Cooperative agreement Grant number DMR-1644779 and the state of Florida.  EJGS acknowledges computational resources through the
UK Materials and Molecular Modelling Hub for access to THOMAS supercluster,
which is partially funded by EPSRC (EP/P020194/1); CIRRUS Tier-2 HPC
Service (ec131 Cirrus Project) at EPCC (http://www.cirrus.ac.uk) funded
by the University of Edinburgh and EPSRC (EP/P020267/1);
ARCHER UK National Supercomputing Service (http://www.archer.ac.uk) via
Project d429.  EJGS acknowledges the
EPSRC Early Career Fellowship (EP/T021578/1) and
the University of Edinburgh for funding support.
The data in this manuscript can be assessed through a request to the corresponding authors.

\setlength{\parindent}{0pt}
\section*{Disclaimer}
The identification of any commercial product or trade name does not imply endorsement or recommendation by the National Institute of Standards and
Technology.

\setlength{\parindent}{0pt}
\subsection*{Competing interests}
The authors declare no competing interests.

\section{Data Availability}\label{Experiments}
Data are available from the corresponding authors upon request.

\setlength{\parindent}{0pt}
\section*{Author contributions}
L.B. conceived the project. J.M. synthesized the Fe$_{3-x}$GeTe$_2$ single-crystals and characterized its magnetic response and electrical transport properties.
D.R., J.M. and Y.C.C. measured the initial anomalous Nernst response. E.S.C. measured anomalous Hall, Nernst and thermal Hall signals on the same crystal via continuous and pulsed current methods. A.W., G.T.M and J.Y.C. characterized via single-crystal x-ray diffraction the initially synthesized Fe$_{3-x}$GeTe$_2$ crystals. G.S.K and M. S. used single-crystal x-ray diffraction to select samples for neutron diffraction measurements. P.P.B., A.J.G., J.A.B. and W.D.R performed neutron scattering measurements. M.S. and E.S performed the calculations and simulations describing the magnetic domains. L.B. wrote the manuscript with input from all authors. All authors discussed the results.

\end{document}